\documentclass[preprintnumbers,superscriptaddress,amsmath,amssymb,nofootinbib,twocolumn,showpacs]{revtex4-1}

\usepackage{dcolumn}
\usepackage{bm}
\usepackage{graphicx}
\usepackage{subcaption}

\usepackage{color}
\usepackage{natbib}
\usepackage{twoopt}
\newcommand{\beq}{\begin{equation}}
\newcommand{\eeq}{\end{equation}}
\newcommand{\ben}{\begin{eqnarray}}
\newcommand{\een}{\end{eqnarray}}
\newcommand{\bi}{\begin{itemize}}
\newcommand{\ei}{\end{itemize}}
\newcommand{\nn}{\nonumber}

\newcommand{\ie}{\textit{i.e.}}
\newcommand{\eg}{\textit{e.g.}}
\newcommand{\etc}{\textit{etc.}}
\newcommand{\etal}{\textit{et al.}}

\newcommand{\citeeq}[1]{Eq.~(\ref{#1})}

\newcommand{\citeeqs}[1]{Eqs.~(\ref{#1})}
\newcommand{\citeeqss}[2]{Eqs.~(\ref{#1})~and~(\ref{#2})}

\newcommand{\citesec}[1]{Sect.~\ref{#1}}

\newcommand{\citetab}[1]{Tab.~\ref{#1}}
\newcommand{\citefig}[1]{Fig.~\ref{#1}}

\newcommand{\msun}{\mbox{$M_{\odot}$}}

\newcommand{\fsol}{{\ifmmode f_{\odot} \else $f_{\odot}$\fi}}

\newcommand{\sqgev}{{\ifmmode {\rm GeV}^2 \else ${\rm GeV}^2$\fi}}

\newcommand{\sr}{{\ifmmode {\rm sr} \else ${\rm sr}$\fi}}
\newcommand{\invsr}{{\ifmmode {\rm sr}^{-1} \else ${\rm sr}^{-1}$\fi}}

\newcommand{\scnd}{{\ifmmode {\rm s} \else ${\rm s}$\fi}}
\newcommand{\invscnd}{{\ifmmode {\rm s}^{-1} \else ${\rm s}^{-1}$\fi}}

\newcommand{\kpc}{{\ifmmode {\rm kpc} \else ${\rm kpc}$\fi}}
\newcommand{\invkpc}{{\ifmmode {\rm kpc}^{-1} \else ${\rm kpc}^{-1}$\fi}}

\newcommand{\sqkpc}{{\ifmmode {\rm kpc}^{2} \else ${\rm kpc}^{2}$\fi}}
\newcommand{\invsqkpc}{{\ifmmode {\rm kpc}^{-2} \else ${\rm kpc}^{-2}$\fi}}

\newcommand{\cm}{{\ifmmode {\rm cm} \else ${\rm cm}$\fi}}
\newcommand{\invcm}{{\ifmmode {\rm cm}^{-1} \else ${\rm cm}^{-1}$\fi}}

\newcommand{\sqcm}{{\ifmmode {\rm cm}^2 \else ${\rm cm}^2$\fi}}
\newcommand{\invsqcm}{{\ifmmode {\rm cm}^{-2} \else ${\rm cm}^{-2}$\fi}}

\newcommand{\meter}{{\ifmmode {\rm m} \else ${\rm m}$\fi}}
\newcommand{\invmeter}{{\ifmmode {\rm m}^{-1} \else ${\rm m}^{-1}$\fi}}

\newcommand{\sqmeter}{{\ifmmode {\rm m}^2 \else ${\rm m}^2$\fi}}
\newcommand{\invsqmeter}{{\ifmmode {\rm m}^{-2} \else ${\rm m}^{-2}$\fi}}

\newcommand{\lcdm}{{\ifmmode \Lambda{\rm CDM} \else $\Lambda{\rm CDM}$\fi}}

\newcommand{\Rvirh}{{\ifmmode R_{\rm vir}^{\rm h} \else 
    $R_{\rm vir}^{\rm h}$\fi}}

\newcommand{\Ncl}{{\ifmmode N_{\rm cl} \else $N_{\rm cl}$\fi}}
\newcommand{\ncl}{{\ifmmode n_{\rm cl} \else $n_{\rm cl}$\fi}}
\newcommand{\phicl}{{\ifmmode \phi_{\rm cl} \else $\phi_{\rm cl}$\fi}}
\newcommand{\phicltot}{{\ifmmode \phi_{\rm cl}^{\rm tot} \else 
    $\phi_{\rm cl}^{\rm tot}$\fi}}
\newcommand{\Beff}{{\ifmmode B_{\rm eff} \else $B_{\rm eff}$\fi}}

\newcommand{\aj}{Astron. J.}
\newcommand{\aap}{Astron. Astroph.}
\newcommand{\apjs}{Astrophys. J. Suppl. Series}
\newcommand{\apjl}{Astrophys. J. Lett.}
\newcommand{\araa}{Ann. Rev. Astron. Astrophys.}

\newcommand{\jcap}{JCAP}
\newcommand{\mnras}{MNRAS}

\newcommand{\physrep}{Phys. Rept.}

\definecolor{myred}{RGB}{102,0,0}

\usepackage[pdftex,
            breaklinks=true,%
            colorlinks=true,%
            linkcolor=myred,
            urlcolor=blue,
            citecolor=blue,
            pdfauthor={Stref, Lavalle},%
            pdftitle={Template for article in XXX}%
           ]{hyperref}

\graphicspath{{./}{Figs/}}

\begin{document}

\title{Modeling dark matter subhalos in a constrained galaxy:\\
  Global mass and boosted annihilation profiles}
\author{Martin Stref}
\email{martin.stref@umontpellier.fr}
\affiliation{Laboratoire Univers \& Particules de Montpellier (LUPM),
  CNRS \& Universit\'e de Montpellier (UMR-5299),
  Place Eug\`ene Bataillon,
  F-34095 Montpellier Cedex 05, France}

\author{Julien Lavalle}
\email{lavalle@in2p3.fr}
\affiliation{Laboratoire Univers \& Particules de Montpellier (LUPM),
  CNRS \& Universit\'e de Montpellier (UMR-5299),
  Place Eug\`ene Bataillon,
  F-34095 Montpellier Cedex 05, France}

\begin{abstract}
  The interaction properties of cold dark matter (CDM) particle candidates, such as those of
  weakly interacting massive particles (WIMPs), generically lead to the structuring of dark matter
  on scales much smaller than typical galaxies, potentially down to $\sim 10^{-10}M_\odot$.
  This clustering translates into a very large population of subhalos in galaxies and affects the
  predictions for direct and indirect dark matter searches (gamma rays and antimatter cosmic rays).
  In this paper, we elaborate on previous analytic works to model the Galactic subhalo population,
  while remaining consistent with current observational dynamical constraints on the Milky Way. In
  particular, we propose a self-consistent method to account for tidal effects induced by both
  dark matter and baryons. Our model does not strongly rely on cosmological simulations, as they
  can hardly be fully matched to the real Milky Way, apart from setting the initial subhalo mass
  fraction. Still, it allows us to recover the main qualitative features of simulated systems. It
  can further be easily adapted to any change in the dynamical constraints, and can be used to
  make predictions or derive constraints on dark matter candidates from indirect or direct
  searches. We compute the annihilation boost factor, including the subhalo-halo cross product. We
  confirm that tidal effects induced by the baryonic components of the Galaxy play a very
  important role, resulting in a local average subhalo mass density $\lesssim 1\%$ of the total
  local dark matter mass density, while selecting the most concentrated objects and leading to
  interesting features in the overall annihilation profile in the case of a sharp subhalo mass
  function. Values of global annihilation boost factors range from $\sim 2$ to $\sim 20$, while
  the local annihilation rate is about half as much boosted.
\end{abstract}

\maketitle
\preprint{LUPM:16-016 hal-01377666}

\tableofcontents

\section{Introduction}
\label{sec:intro}
While the long-standing issue of the origin of dark matter (DM) is still pending, many experiments 
involved in this quest have recently reached the sensitivity to probe the relevant parameter space
for one of the most popular particle candidates, the WIMP, which finds specific realizations in
many particle physics scenarios beyond the standard model 
(\eg~Refs.~\cite{Primack1988,Jungman1996,Bergstroem2000}).
Among different search strategies, indirect DM searches 
(\eg~Refs.~\cite{Carr2006,Porter2011,Lavalle2012a,Strigari2013}) are 
becoming quite constraining for WIMPs annihilating through s-waves. This is particularly 
striking not only for indirect searches in gamma rays
(\eg~Refs.~\cite{Gunn1978,Bergstroem1988,Conrad2015a}), 
but also in the antimatter cosmic-ray spectrum~\cite{Silk1984}, both with positrons 
(\eg~Ref.~\cite{Bergstroem2013}) and with antiprotons (\eg~Ref.~\cite{Giesen2015}). For indirect
searches, the way the Galactic dark matter halo is modeled is a fundamental piece in deriving
constraints or testing detectability. For direct DM searches, whether the local DM density is
smooth or may contain inhomogeneities also has important consequences
(see \eg~Ref.~\cite{Freese2013}).

A generic cosmological consequence of the WIMP scenario (among other CDM candidates) is the 
clustering of dark matter on very small, subgalactic scales, when the Universe enters the
matter-domination era
(\eg~Refs.~\cite{Peebles1984,Silk1993,Kolb1994a,Chen2001,Hofmann2001,Berezinsky2003,Green2004a,Loeb2005,Boehm2005a,Profumo2006,Bertschinger2006a,Bringmann2007,Visinelli2015}, and Ref.~\cite{Bringmann2009} for a review). Both analytic calculations 
(see a review in \eg~Ref.~\cite{Berezinsky2014}) and cosmological 
simulations (\eg~Refs.~\cite{Diemand2005a,Diemand2007a,Springel2008,Anderhalden2013,Ishiyama2014})
show that many of these subhalos survive in galaxies against tidal disruption, and
further constrain their properties. Consequently, the DM halo embedding the MW, if it is made of
WIMPs, is not a smooth distribution of DM, but instead exhibits inhomogeneities in the form of many
subhalos or their debris. In the context of 
self-annihilating DM candidates, this leads to the interesting consequence of enhancing the 
average annihilation rate with respect to the smooth-halo assumption \cite{Silk1993}. Generic 
methods to account for a subhalo population in the DM annihilation signal predictions were 
originally presented in Refs.~\cite{Bergstroem1999a,Ullio2002a} for gamma rays, and in 
Refs.~\cite{Lavalle2007a,Lavalle2008c} for antimatter cosmic rays.

While subhalos are now very often included when deriving constraints from the Galactic or 
extragalactic diffuse gamma-ray emissions (see
\eg~Refs.~\cite{Ullio2002a,Pieri2011,Blanchet2012,Serpico2012,Ajello2015}, and 
a review in Ref.~\cite{Fornasa2015}), this is still barely the case for the antimatter 
channels (\eg~Refs.~\cite{Giesen2015,Boudaud2015}). In the latter case, although it was shown
that subhalos could not enhance the predictions by orders of magnitude
\cite{Lavalle2008c,Pieri2011}, the precision achieved by current experiments
(see \eg~Refs.~\cite{Adriani2009a,Adriani2010a,Adriani2014a,AguilarAliCavasonzaAlpatEtAl2016} for
antiproton measurements) implies that even small changes in the predicted fluxes could still have
a strong impact on constraints on the WIMP mass. In this paper, our aim is to provide a dynamically
self-consistent model of a Galactic subhalo component in order to improve the constraints derived
on s-wave annihilating WIMPs.

The paper develops as follows: A short overview of our method is presented in \citesec{sec:sketch},
where the main steps are made clear. The instrumental part of our study is described in
\citesec{sec:dmhalo}, where we introduce the dark halo setup including both a smooth and a subhalo
component, and where we discuss the tidal effects induced by both baryons and dark matter.
We then discuss the mass profiles, the dark matter annihilation profile, and the corresponding
differential and integrated annihilation boost factors in \citesec{sec:res}, which can be used in
indirect detection studies. In that part, we also quantify the theoretical uncertainties 
coming from using different Galactic mass models, different tidal cutoff criteria, or other
subhalo population properties. We conclude and present our perspectives in \citesec{sec:concl}.
\section{Overview of the model}
\label{sec:sketch}

We summarize below the main steps of the procedure we have developed in this study to
  get both a smooth Galactic halo and its subhalo population consistent with dynamical constraints:
  \bi
\item[1.] Select a complete and constrained Milky Way mass model that includes baryons (disk,
  bulge, {\em etc.}) and a DM halo, and which provides a good fit to kinematic data.
\item[2.] Assume that the DM halo is separable in terms of a smooth component and a subhalo
  components which should be true from the initial stage of the Galactic halo formation to its
  currently observed state [see \citesec{ssec:halo_and_sub} and \citeeq{eq:rhotot}] -- during the
  evolution of the halo, part of the subhalo component mass is stripped away by tidal effects and
  will be considered as fully transferred to the smooth component. Assume that equilibrium has been
  reached today.
\item[3.] Assign a phase space to the subhalo component: each subhalo is considered to be an
  independent object characterized by (i) its initial mass (in a homogeneous background density),
  (ii) its concentration, and (iii) its position in the Milky Way (spherical orbits assumed) --
  associated distribution functions are initially independent (factorizable), but will become
  intricate (non-factorizable) as gravitational tidal effects come into play
  [see \citesec{sssec:overall_sub} and item 5 below].
\item[4.] Assume that the concentration and mass functions are initially the
  cosmological ones, \ie\ position-independent [see \citesec{sssec:dPdc} and
    \citesec{sssec:dPdm}]. Further assume that both the smooth and the subhalo components were
  initially following the same spatial distribution, which sets the initial spatial distribution
  for subhalos before tidal stripping and potential subsequent disruption
  [see \citesec{sssec:dPdV}].
\item[5.] Normalize the whole subhalo mass (or the total number of subhalos) from preferred
  prescriptions [this can be done from observational or structure formation constraints; before or
    after plugging tidal stripping, \ie\ after item 6a or 6b below; see
    \citesec{sssec:norm_sub}].
\item[6.] Determine how gravitational tides due to both the baryonic components and the whole dark
  matter content affect subhalos as a function of their properties (location, mass, and
  concentration), which allows to get the final phase-space distribution for subhalos
  [\citesec{ssec:tides} on tidal
    effects, which is a crucial part in this work]. In practice, we proceed by:
  \bi
\item[a.] calculating the global tidal stripping induced by the host halo, which sets a first
  tidal radius $r_t$ [see \citesec{sssec:global_tides}];
\item[b.] calculating the tidal stripping induced by disk shocking, which may reduce
  the tidal radius initially set by global halo tides [strong reduction in the central parts of
    the Galaxy, see \citesec{sssec:disk_tides}];
\item[c.] determining a criterion for disrupting subhalos depending on their tidal radii --
  so-called tidal disruption efficiency $\varepsilon_t$ [see \citesec{sssec:disruption}].
  \ei
  The obtained final intricate phase-space function is mostly determined by the new concentration
  function, which has become spatial-dependent as a result of tidal disruption. This implies that
  the mass function also becomes spatial-dependent since integrated over concentration, and the
  initial spatial distribution gets modified, since integrated over mass. This final
  phase-space distribution should provide a more realistic description of subhalos, including tidal
  stripping and disruption inferred from all components of the constrained Galactic mass model,
  and therefore consistent with dynamical constraints by construction [see again
    \citesec{sssec:overall_sub} for the description of the intricate phase space].
  It is encoded in \citeeqss{eq:def_nsub}{eq:def_rhosub} which, together with \citeeq{eq:rhotot},
  define our constrained global DM halo model.
\item[7.] Use this final subhalo distribution to get the smooth component from the overall
  constrained halo profile [see \citeeq{eq:rhotot}], and to further compute observables relevant
  to DM searches [\eg\ local density of subhalos, annihilation boost factor, \etc\ -- see
    \citesec{sec:res}].
\ei
Not only this theoretical modeling allows to recover the main qualitative
results obtained in zoomed cosmological simulations (\eg\ the fact that the whole subhalo
distribution is strongly depleted in the central regions of the Galaxy, while it dominates the
mass profile in the outskirts), but it also allows to get a complete DM halo (including a smooth
component and a subhalo component) which is fully consistent with current dynamical constraints.
The latter point can actually hardly be achieved when directly importing results from
cosmological simulations. Indeed, the dark and baryonic profiles found in Milky Way-like
simulations, even if somewhat similar, can barely be fully matched to the detailed observed
properties of the Milky Way, which are strongly constrained by kinematic data. Since these
detailed properties play a central role in terms of tidal stripping, blindly extrapolating
results from cosmological simulations will likely be plagued by inconsistencies, which we remedy
here in a theoretically consistent, reproducible, and tuneable way. Instead, simulations remain
instrumental to get insight on the physical processes themselves.

\section{The Milky Way dark halo and its subhalo population}
\label{sec:dmhalo}
In this section, we propose a self-consistent method to constrain the subhalo population of the 
MW dark halo and to derive therein the DM annihilation rate including all components. This 
method subscribes to two main principles: (i) accounting for existing dynamical constraints in the 
MW; (ii) starting from general assumptions, then comparing to and calibrating on high-resolution 
cosmological simulations only {\em a posteriori}.

In the following, any halo mass $m$ will, unless
specified otherwise, express the mass contained within a sphere of radius $r_{200}$ such that
\ben
\label{eq:m200}
m = m(r_{200}) = m_{200} = \frac{4\,\pi}{3}\,(200\times\rho_c)\,r_{200}^3\,,
\een
where $\rho_c$ is the critical density of the Universe as measured today, which we compute from
the best-fit Hubble parameter obtained by the Planck Collaboration (combined analysis):
$H_0=67.74$ km/s/Mpc.
\subsection{Dark halo model}
\label{ssec:halo_and_sub}
The most basic and obvious assumption one can make about the DM distribution in the Galaxy is
that the DM density profile $\rho_{\rm tot}$ can be split into two components: one smooth, 
$\rho_{\rm sm}$, and another made of subhalos, $\rho_{\rm sub}$, such that at any position $\vec{x}$,
\ben
\label{eq:rhotot}
\rho_{\rm tot}(\vec{x}) = \rho_{\rm sm}(\vec{x}) + \rho_{\rm sub}(\vec{x})\,,
\een
such that the total dark mass is given by
\ben
\label{eq:M200}
M_{200} = \int_{V_{200}} dV\,\rho_{\rm tot}(\vec{x})\,,
\een
where $V_{200}$ is the spherical volume delineated by the associated pseudovirial radius $R_{200}$.

Furthermore, to get reliable predictions for DM annihilation signals, it is important to account
for existing dynamical constraints on the DM profile. There have been significant efforts
to improve Milky Way mass models in the recent years (\eg~Refs.~\cite{Catena2010a,Salucci2010,McMillan2011,Bovy2012b,Bovy2013,Piffl2014,Kafle2014,Bienayme2014,McKee2015,Xia2016,Huang2016}), such that
the dark halo is actually strongly constrained in its shape and related parameters.
Modeling a clumpy dark halo in the context of DM searches can therefore strongly benefit from
these results, and in any case should account for existing dynamical constraints. We stress that
global dynamical studies like those cited above provide constraints on  $\rho_{\rm tot}$, but not
on $\rho_{\rm sm}$ and $\rho_{\rm sub}$ separately.

In many Galactic DM search studies involving subhalos, one usually exploits the results of
high-resolution cosmological simulations either by putting a virtual observer at 8 kpc from the
center of the simulated halo (irrespective of the differences between the real and simulated
galaxy) and computing relevant observables, by rescaling simulation profiles to match with the
measured local DM density, or by adding simulation-inspired fits of subhalo number density profiles
to get predictions. However, as we will show later, tidal effects have a strong dependence on the
details of the (baryonic and dark) matter content of the Milky Way. Therefore, since the Milky Way
halo and its baryonic content are now rather strongly constrained, such blind matchings or
extrapolations are likely to provide uncontrolled, or at least inconsistent, results
(even if dubbed Milky Way-like, a cosmological simulation can hardly be fully matched to the
Milky Way -- \eg\ the detailed halo profile and its parameters; the precise size, width, mass
of the disk; intrinsic mass resolution limit; \etc\ --  but to some extent). This motivates us
to go beyond these simplistic recipes and propose a new approach. We still emphasize that
cosmological simulations do provide very important and useful information about the subhalo
dynamics, and we will take advantage of the {\em generic} subhalo properties inferred from
simulations rather than the {\em peculiar} description of a single simulated clumpy halo, even
though the latter case provides a very nice environment to study dynamical correlations between
various galactic components or more specific physical processes.

From cosmological structure formation (see \eg~Refs.~\cite{Cooray2002,Zentner2007,Knobel2012}), we
know that galactic halos form rather late ($z\sim 6$) with respect to the smallest-scale halos
expected in the WIMP scenario ($z\sim 80$). It is therefore reasonable to assume that the smooth
and subhalo components follow the same spatial distribution when the Galactic halo forms.
Then, as the Galaxy evolves, several changes occur: (i) further subhalos are accreted; and (ii)
subhalos may experience mergers, stellar encounters, and tidal disruptions. Since the former 
phenomenon also concerns the smooth component, it should not modify the overall picture
(subhalos may be considered as test particles among others). However, 
the latter must be taken into account, since it will reduce the subhalo number density in regions 
close to the terrestrial observers. This approximate trend is actually what is found in very 
high-resolution cosmological simulations, where the subhalo number density is shown to depart from 
the overall DM distribution essentially in the central regions of 
galaxies~\cite{Diemand2008b,Springel2008,Zhu2016}. In the same references, the global DM profile
(including subhalos) is found to be consistent with the seminal 
earlier results obtained by Navarro, Frenk, and White~\cite{Navarro1996} (hereafter NFW) and 
subsequent refinements (\eg~Refs.~\cite{Navarro2004,Merritt2006a,Diemand2008b,Navarro2010,Zhu2016}).
Inner cored profiles can also be found as a result of efficient feedback originating in
star formation and supernova explosions \cite{Maccio2012a,Mollitor2015}.

All this suggests the following method to try to build a self-consistent dark halo with a 
substructure component: (i) assume a global DM halo profile $\rho_{\rm tot}$ constrained by dynamical
studies; (ii) start with a subhalo population tracking the smooth halo, such that both 
$\rho_{\rm sub}\propto \rho_{\rm tot}$ and $ \rho_{\rm sm}\propto \rho_{\rm tot}$; (iii) plug in tidal 
disruption such that the mass contained in disrupted subhalos and in the pruned part of the 
survivors is transferred to the smooth-halo component; (iii) compare/cross-calibrate the final 
result with/onto high-resolution cosmological simulations. Before we translate this method in
terms of equations for DM searches, we need to figure out how to express the mass density profile 
$\rho_{\rm sub}$ associated with subhalos. In practice, the smooth DM component will merely be 
determined from \citeeq{eq:rhotot} as $\rho_{\rm sm}=\rho_{\rm tot}-\rho_{\rm sub}$, after having set 
$\rho_{\rm sub}$.
\subsection{Accounting for dynamical constraints}
\label{ssec:dynamics}
As a template and dynamically constrained global dark halo, we will use the best-fit MW mass model
obtained by McMillan~\cite{McMillan2011} (M11 hereafter), which turns out to be fully consistent
with more recent studies (\eg~Refs.~\cite{Piffl2014,Kafle2014,McMillan2016}) while rather simple to
implement. This model was derived from a Bayesian analysis run upon several observational data sets,
photometric as well as kinematic, restricting to the terminal velocity curves measured for
longitudes $|l|>45^\circ$ --- this model does not address the complex structure of the very central
regions of the MW, nor does it include any atomic or molecular gas component (we will use mass
models including gas components in \citesec{ssec:comp_mass_models}).
\subsubsection{Global dark halo and baryons}
\label{sssec:darkhalo}
M11 assumes a spherically symmetric NFW profile, given in terms of the general $\alpha\beta\gamma$
parametrization~\cite{Hernquist1990,Zhao1996} as
\ben
\rho_{\rm tot}(r) = \rho_{\alpha\beta\gamma}(r)
\equiv \rho_s\,\left(r/r_s\right)^{-\gamma}\,
\left\{ 1+ \left(r/r_s\right)^\alpha \right\}^{-\frac{(\beta-\gamma)}{\alpha}}\,,\nn\\
\label{eq:dmprofile}
\een
with $(\alpha,\beta,\gamma)=(1,3,1)$ for an NFW profile. The M11 best-fit values for the scale
density $\rho_s$ and the scale radius $r_s$ are given in
\citetab{tab:dark_halo}.
For the sake of comparison, we also introduce the Einasto dark matter profile
\cite{Einasto1965,Navarro2004}:
\ben
\rho_{\rm ein}(r) = \rho_s \,
\exp\left\{ -\frac{2}{\alpha_e}\left[ \left(  \frac{r}{r_s} \right)^{\alpha_e} -1 \right] \right\}\,.
\een
This profile halo was used in a dynamical study complementary to and consistent with M11, presented
in Ref.~\cite{Catena2010a} (CU10 hereafter). The associated parameters are also given in
\citetab{tab:dark_halo}. Irrespective of the MW mass model, these dark matter profiles will also
be used to describe the inner density profiles of subhalos (before tidal
  stripping). In the following, we will use M11 as our reference case.

Since we also aim at considering the baryonic components when dealing with tidal effects (see 
Ref.~\cite{Bland-Hawthorn2016} for a recent review), we provide the axisymmetric M11 bulge-disk 
density model below (with the convention $r^2=R^2+z^2$), where the subscript $b$ refers to the bulge
and $d$ to the disk:
\ben
\rho_b(R,z) &=& \frac{\rho_b}{(1+r'/r_b)^{\alpha_b}}\,
\exp\left\{ -\left( \frac{r'}{r_b^c}\right)^2\right\}\;,\nn\\
\rho_d(R,z) &=& \frac{\Sigma_d}{2\,z_d}\,
\exp\left\{ - \frac{R}{R_d} - \frac{|z|}{z_d} \right\}\;,
\label{eq:rho_baryons}
\een
where $r'\equiv \sqrt{R^2+(z/q)^2)}$, $q$ is the axial ratio, $\Sigma_d$ is the disk surface
density, and the other parameters are scale parameters. All parameters are given in 
\citetab{tab:baryons}, where a two-component disk is explicit (thin and thick disks) -- note
that the above disk parameterization can also be relevant to additional gas components
(see \citesec{ssec:comp_mass_models}). Since the
model was not fitted against observational data featuring the central regions of the Galaxy,
the bulge parameters but $\rho_b$ are actually fixed to those obtained in Ref.~\cite{Bissantz2002}.
Note that such a disk profile can also be relevant to describing gaseous components, which have not
been included in M11.

It will prove useful to have a spherical approximation of the disk density when dealing with
global tides (see \citesec{sssec:global_tides}). We readily derive it by demanding that the disk
mass inside a sphere of radius $r$ equal the actual disk mass inside an infinite cylinder of
radius $R$. It reads
\ben
\rho_{d,{\rm sph}}(r) &=& \frac{\Sigma_d}{2\,r}\,
\exp\left\{ - \frac{r}{R_d}\right\}\;.
\label{eq:sph_disk}
\een
One may find similar expressions with $R_d\leftrightarrow\sqrt{R_d^2+z_d^2}$
(\eg\ Ref.~\cite{Binney2008}), but using one or another has absolutely no impact in this study.
\begin{table*}[t]
  \begin{tabular}{| c || c | c | c | c | c | c | c |}
    \hline
    MW mass model & Profile & $r_{200}$ & $M_{200}$ & $r_s$ & $\rho_s$ & $R_\odot$ &  $\rho_\odot$  \\
    & & [kpc] & [$M_\odot$] & [kpc] & [GeV/cm$^3$] & [kpc] & [GeV/cm$^3$] \\
    \hline
    M11 & NFW & 237 & $1.43\times 10^{12}$ & 20.2 & 0.32 & 8.29 & 0.395 \\
    \hline
    CU10 & Einasto($\alpha_e=0.22$) & 208 & $9.6\times 10^{11}$ & 16.07 & 0.11 & 8.25 & 0.386 \\
    \hline
    M16 & NFW & 230.5 & $1.31\times 10^{12}$ & 19.6 & 0.32 & 8.21 & 0.383 \\
    \hline
  \end{tabular}
  \caption{\small Dark matter halo parameters for different Galactic mass models (best-fit models of
    Refs.~\cite{McMillan2011} [M11], \cite{Catena2010a} [CU10], and \cite{McMillan2016} [M16]).}
  \label{tab:dark_halo}
\end{table*}

\begin{table*}[t]
  \begin{tabular}{| c || c | c | c | c | c | c | c | c | c | c | c | c | c |}
    \hline
    MW mass model  & $q$ & $\alpha_b$ & $r_b$ & $r_b^c$ & $\rho_b$ & $R_d$ [kpc] & 
    $z_d$  [kpc] & $\Sigma_d$ [$M_\odot/{\rm pc}^2$] \\
    & & & [kpc] & [kpc] & [$M_\odot/{\rm pc}^3$] & (thin/thick)(HI/HII) & (thin/thick)(HI/HII) &
    (thin/thick)(HI/HII) \\
    \hline
    M11 & 0.5 & 1.8 & 0.075 & 2.1 & 95.6 & (2.9/3.31)(-/-) & (0.3/0.9)(-/-) & (816.6/209.5)(-/-) \\
    \hline
    CU10 & 0.6 & 1.85 & 0.3879 & 0.872 & 1.37 & (2.45/-)(7/1.5)$^\dagger$ & (0.34/-)(0.085/0.045)$^\dagger$ & (1154.12/-)(53.1/2180)$^\dagger$ \\
    \hline
    M16 & 0.5 & 1.8 & 0.075 & 2.1 & 98.4 & (2.5/3.02)(7/1.5) & (0.3/0.9)(0.085/0.045) & (896/183)(53.1/2180)\\
    \hline
  \end{tabular}
  \caption{\small Baryonic component parameters for different Galactic mass models (best-fit
    models of Refs.~\cite{McMillan2011} [M11], \cite{Catena2010a} [CU10], and \cite{McMillan2016}
    [M16]). $^\dagger$: The CU10 HI and HII gas disks are inferred from old data points, so we adopt
    the same parameterization as in M16 for simplicity -- this has negligible impact on the final
    results.}
  \label{tab:baryons}
\end{table*}
\subsubsection{The overall subhalo component}
\label{sssec:overall_sub}
The very presence of subhalos in the Galactic host halo leads to strong DM inhomogeneities,
so defining a global regular mass density function for subhalos implicitly implies averaging over a 
certain volume. In the following, we will assume that the whole Galaxy has reached an
equilibrium state (no time dependence), and that subhalos are independent objects described
over a phase space $w^n$ that includes their position $\vec{x}$, mass $m$, and concentration $c$
(we define these parameters in \citesec{ssec:sub}). We also assume that a subhalo at position
$\vec{x}$ can be described as an individual object lying in a local background density
$\rho_{\rm tot}(\vec{x})$ constant over the size of the subhalo (coarse-grain approximation). With
these assumptions, the subhalo phase-space number density reads
\ben
\label{eq:subps}
\frac{d^n N}{dw^n} &=& \frac{N_{\rm sub}}{K_w} \, 
\frac{d{\cal P}_V(\vec{x})}{dV}\,
\frac{d{\cal P}_m(m)}{dm}\,
\frac{d{\cal P}_c(c,m,\vec{x})}{dc}\;\nn\\
&=& \frac{N_{\rm sub}}{K_w} \, \frac{d{\cal P}_{V,m,c}(\vec{x},m,c)}{dV\,dm\,dc}\,.
\een
Without gravitational interactions between subhalos and the rest of the Milky Way,
the phase space would be separable such that each ${\cal P}_w$ could be factorized. However,
as we will see later, tidal effects induce non-trivial correlations between $\vec{x}$, $m$,
and $c$, and the phase space becomes intricate. However, the individual ${\cal P}_w$s still
define here effective ``initial'' conditions before tidal effects are plugged in, and are also
necessary inputs to calculate the final {\em local} phase space (intricate concentration and
mass functions at a given position $\vec{x}$).
Parameter $K_w$ featured above is a normalization constant determined by the following closure
relation:
\ben
\int dw^n\, \frac{d^n {\cal P}_w}{dw^n}&=& K_w\nn\\
\Leftrightarrow \int dw^n\, \frac{d^n N}{dw^n}&=& N_{\rm sub}\;,
\label{eq:subnorm}
\een
where $N_{\rm sub}$ is the total number of subhalos over the whole phase space embedded in 
the host dark halo. Each individual probability distribution function (PDF) $d{\cal P}_{w_i}/dw_i$, 
where $w=V,m,c$ ($V$ is the physical volume), is defined such that it is normalized over its 
phase-space subvolume $\delta W_i$ as
\ben
\int_{w_{i,{\rm min}}\in \delta W_i}^{w_{i,{\rm max}}\in \delta W_i} dw_i\, \frac{d {\cal P}_{w_i}}{dw_i} = 1\,.
\een
We emphasize that as long as these individual PDFs are uncorrelated (``initial''
conditions), $K_w=1$, but this is generally not the case. In particular, tidal effects
imply that each subhalo is actually featured by a tidal radius $r_t$ which depends on the
initial subhalo mass $m$, its position $\vec{x}$ in the Galactic halo, and its concentration $c$
--- we will detail the individual PDFs in \citesec{ssec:sub} and discuss the tidal disruption of
subhalos in \citesec{ssec:tides}. Therefore, tidal effects do induce an explicit correlation
between the PDFs, making the subhalo phase space intricate and non-trivial, and leading here to
$K_w \neq 1$.
\begin{widetext}
  However, we can still self-consistently define the global subhalo number density $n_{\rm sub}$
  profile as
    \ben
    \frac{dn_{\rm sub}(\vec{x},m)}{dm} &=&
    \frac{N_{\rm sub}}{K_w} \, \frac{d{\cal P}_V(\vec{x})}{dV}\,
         \frac{d{\cal P}_m(m)}{dm}
         \int_{c_{\rm min}(\vec{x})}^{c_{\rm max}} dc \,\frac{d{\cal P}_c}{dc}\nn\\
         \Longrightarrow \;\; 
         n_{\rm sub}(\vec{x}) &=&
         \frac{N_{\rm sub}}{K_w} \, \frac{d{\cal P}_V(\vec{x})}{dV}
         \int_{m_{\rm min}}^{m_{\rm max}}dm\,\frac{d{\cal P}_m(m)}{dm}
         \int_{c_{\rm min}(\vec{x})}^{c_{\rm max}} dc \,\frac{d{\cal P}_c}{dc}\,,
\label{eq:def_nsub}
\een
where we wrote both the mass-differential form and its integral. The differential
form is here expressed in terms of $m$, \ie\ the initial (cosmological) subhalo mass
contained inside an approximate virial radius assuming a homogeneous background matter,
usually $r_{200}$, which is not the actual tidal subhalo mass $m_t$ (see \citesec{ssec:sub} for
details) -- we will show how to obtain
the differential form in terms of $m_t$ in \citesec{ssec:nsub}. It will also become
clear in \citesec{sssec:disruption} why tidal effects induce a spatial dependence of the minimal
allowed concentration $c_{\rm min}(\vec{x})$, such that $n_{\rm sub}$ is actually very strongly
depleted toward the central regions of the Galaxy in spite of the increase of $d{\cal P}_V/dV$.

We can also self-consistently define the global subhalo mass density profile as
\ben
\rho_{\rm sub}(\vec{x}) &=& N_{\rm sub}\,\widetilde{\langle m_t\rangle}(\vec{x})\,
\frac{d{\cal P}_V(\vec{x})}{dV}\,,\nn\\
{\rm with} \; \widetilde{\langle m_t \rangle}(\vec{x})&\equiv& 
\frac{1}{K_w} \,
\int_{m_{\rm min}}^{m_{\rm max}} dm \,\frac{d{\cal P}_m}{dm}
\int_{c_{\rm min}(\vec{x})}^{c_{\rm max}} dc \,\frac{d{\cal P}_c}{dc}\,
m_t(r_t(c,m,\vec{x}),m,c)\;,
\label{eq:def_rhosub}
\een
where $m_t$ is the tidal subhalo mass contained within the tidal radius $r_t$, to be contrasted
again with $m$. The symbol $\widetilde{\langle \rangle}$ is {\em not} the average over the
mass and concentration subpart of the phase space because of the normalization $K_w$,
which is calculated over the full phase space. The real mean mass (or any other quantity
depending on mass and concentration) is actually given by
\ben
\langle m_t \rangle (\vec{x}) =
\frac{ \int_{m_{\rm min}}^{m_{\rm max}} dm \,\frac{d{\cal P}_m}{dm}
  \int_{c_{\rm min}(\vec{x})}^{c_{\rm max}} dc \,\frac{d{\cal P}_c}{dc}\,
  m_t(r_t(c,m,\vec{x}),m,c) }{\int_{m_{\rm min}}^{m_{\rm max}} dm \,\frac{d{\cal P}_m}{dm}
  \int_{c_{\rm min}(\vec{x})}^{c_{\rm max}} dc \,\frac{d{\cal P}_c}{dc} }\,.
\label{eq:mmean}
\een
The dependence of the tidal radius $r_t$ on position, mass, and concentration will be discussed
in \citesec{ssec:tides}. Notice that there is also spatial dependence hidden in the denominator
above, as the minimal concentration will be shown to be spatial dependent in
\citesec{sssec:disruption}.

The total mass $M_{\rm sub}^{\rm tot}$ in the form of subhalos is thereby given by
\ben
\label{eq:msubtot}
M_{\rm sub}^{\rm tot} = N_{\rm sub} \, \int_{\rm host \,halo} dV \,\widetilde{\langle m_t\rangle}(\vec{x})
\,\frac{d{\cal P}_V(\vec{x})}{dV}\,.
\een
It will also prove useful to define the total subhalo mass contained in a specific subhalo mass 
subrange $\Delta_{12}^m = [m_1,m_2]\subset[m_{\rm min},m_{\rm max}]$,
\ben
\label{eq:msubref}
M_{\rm sub}^{\Delta_{12}^m} = N_{\rm sub} \, \int_{\rm host \,halo} dV \,
\widetilde{\langle m_t \rangle}_{\Delta_{12}^m}(\vec{x}) \,\frac{d{\cal P}_V(\vec{x})}{dV}\,,
\een
with
\ben
\label{eq:mmeantilde}
\widetilde{\langle m_t\rangle}_{\Delta_{12}^m}(\vec{x})&\equiv& 
\frac{1}{K_w} \,
\int_{m_1}^{m_2} dm 
\int_{c_{\rm min}}^{c_{\rm max}} dc \,
m_t(r_t(c,m,\vec{x}),m,c)\,\frac{d{\cal P}_m}{dm}\,\frac{d{\cal P}_c}{dc}\;.
\een
\end{widetext}
From \citeeqs{eq:M200} and (\ref{eq:msubref}), we can then define the total dark mass fraction in
the form of subhalos within the mass range $\Delta_{12}^m$,
\ben
\label{eq:subfrac}
f_{\rm sub}^{\Delta_{12}^m}\equiv \frac{M_{\rm sub}^{\Delta_{12}^m}}{M_{200}}\,.
\een
We will actually use this fraction to normalize our subhalo population and to calculate 
$N_{\rm sub}$, which we discuss in the next section.
\subsubsection{Calibration of the subhalo component}
\label{sssec:norm_sub}
The overall subhalo distribution being defined, we need to calibrate the subhalo mass content.
To proceed, we will first rely on cosmological simulation results, which provide pictures of MW-like
halos at redshift $z=0$, with subhalo populations that have already experienced all relevant 
dark-matter-only nonlinear disruption or pruning processes 
(see \eg~Refs.~\cite{Zentner2003,Hayashi2003,Diemand2008b,Springel2008,Zhu2016}).
Calibration from first principles is also possible, while more involved and subject to large
theoretical uncertainties; this gives similar constraints though, as reviewed in 
Ref.~\cite{Berezinsky2014}. Besides, it is well known that cosmological parameters have significant 
impact on the global and structural properties of subhalos, especially the matter abundance 
$\Omega_m$, the normalization of the power spectrum $\sigma_8$, and the inflation spectral index 
$n_s$ (see \eg~Refs.~\cite{Press1974,Bond1991a,Lacey1993,Sheth1999} and 
Refs.~\cite{Zentner2003,Dutton2014,Dooley2014}). For instance, larger values of the former
lead to more concentrated halos on all scales, while larger values of the latter increases the power
on small scales -- not to mention the strong correlations between these parameters, such that
an increased $\Omega_m$ can be compensated by a decreased $\sigma_8$ without leading to
significant differences in terms of abundance and clustering properties \cite{GuoEtAl2013}.
One should still favor references with input cosmological parameters not too far from the most 
recent estimates. In particular, the Planck mission 
\cite{PlanckCollaboration2015p} has provided combined constraints, $\Omega_m\simeq 0.31$, 
$\sigma_8\simeq0.82$, and $n_s\simeq 0.97$, directly relevant to the structuring of DM 
subhalos --- note, though, that there are still mild tensions between different cosmological probes 
(see \eg~Ref.~\cite{Riess2016} for a recent illustration). This 
makes the Via Lactea II ultrahigh-resolution simulation \cite{Diemand2008b} (VL2 hereafter) a 
rather conservative reference, since it was run with WMAP-3 best-fit parameters, 
$\Omega_m\simeq 0.24$, $\sigma_8=0.74$, and $n_s=0.951$ \cite{Spergel2007}. For comparison,
the Aquarius simulation series~\cite{Springel2008} were run with $\Omega_m = 0.25$, $\sigma_8=0.9$, 
$n_s=1$, and with a spatial resolution similar to VL2.

We will use the VL2 results to calibrate the subhalo mass fraction defined in 
\citeeq{eq:subfrac}, but the method presented below can be used with any calibration source.
In particular, the authors of VL2 provide the cumulative number of subhalos 
$N_{\rm VL2}(>v_{\rm max})$ as a function of the maximal velocity $v_{\rm max}$. Note that their census
is made up to the host halo radius $R_{50}$ and not $R_{200}$, as defined in \citeeq{eq:m200}. A 
very good fit to this measurement is obtained in the range 
$v_{\rm max}\in[3\,{\rm km/s},20\,{\rm km/s}]=
[v_{\rm max,1},v_{\rm max,2}]$ with the following parametrization \cite{Diemand2008b}:
\ben
N_{\rm VL2}(>v_{\rm max}) = 0.036\left(\frac{v_{\rm max}}{v_{\rm max,host}}\right)^3\,,
\een
with $v_{\rm max,host}=201$ km/s as the maximal velocity of the host halo. The maximal velocity
is directly measured in simulations, and is related to the (sub)halo profile through the relation
\ben
v_{\rm max} = \max\left(\sqrt{\frac{G\,M(r)}{r}}\right)=\sqrt{\frac{G\,M(r_{v_{\rm max}})}{r_{v_{\rm max}}}}\,,
\een
which defines the radius $r_{v_{\rm max}}$, and which can easily be computed for any choice of
subhalo profile once its parameters are fixed (total mass $m$, concentration $c$, and position
$\vec{x}$ in a host halo, if relevant). We can therefore calculate the {\em effective} mass
fraction contained in subhalos within the mass range $[m_1=m(v_{\rm max,1}), m_2=m(v_{\rm max,2})]$ as
\ben
\tilde{f}_{\rm sub,VL2}^{\Delta_{12}^m} &=& \frac{M_{\rm sub,VL2}^{\Delta_{12}^m}}{M_{200}^{\rm VL2}}\\
{\rm with}\; 
M_{\rm sub,VL2}^{\Delta_{12}^m}&\equiv& 
\int_{v_{\rm max,1}}^{v_{\rm max,2}} dv_{\rm max}\,m(v_{\rm max})\,\frac{dN_{\rm VL2}}{dv_{\rm max}}\,.\nn
\een 
This is an {\em effective} mass fraction since it is computed from the pseudovirial mass $m=m_{200}$
instead of the tidal mass $m_t$ used in \citeeqss{eq:msubref}{eq:subfrac}, which is unknown
here. The subtlety is that the subhalo population under scrutiny here has actually already 
experienced tidal effects, which we will ultimately have to account for.
However, at this stage, this effective mass fraction is very useful because its derivation
does not rely on any tidal stripping calculation. Tidal stripping effects will only come
into play at the normalization stage [see discussion after \citeeq{eq:normsub}].

Assuming that VL2 subhalos are well fitted by NFW profiles, and taking a concentration 
function matching the VL2 results (we actually take the VL2 concentration fit proposed in 
Ref.~\cite{Pieri2011}), we find that the total effective subhalo mass in the mass range 
$[m(v_{\rm max,1})=3.14\times 10^6\,\msun,m(v_{\rm max,2})=1.25\times 10^9\,\msun]$ is 
$M_{\rm sub,VL2}^{\Delta_{12}^m} = 2.24\times 10^{11}\,\msun$. Taking the global VL2 halo mass 
$M_{50}=1.93\times 10^{12}\msun$, we obtain an effective mass fraction of $\tilde{f}_{50} = 11.6\%$.
If we further assume that the subhalo number density profile spatially tracks the global halo 
density profile in the outer halo regions, then the extrapolation to $R_{200}<R_{50}$ is trivial:
$\tilde{f}_{200} \simeq \tilde{f}_{50}$. For completeness, we can express this result in terms
of a relative mass range, as different halo models come with different global masses. From the
Einasto profile fitted on the VL2 host halo (see the caption of Fig.~2 in Ref.~\cite{Diemand2008b}),
we get $M_{200}^{\rm VL2}=M^{\rm VL2}(R_{200}^{\rm VL2}=225.44\,{\rm kpc}) = 1.42\times 10^{12}\msun$.
This allows us to propose the following ansatz to normalize the subhalo population:
\ben
\label{eq:normsub}
\tilde{f}_{\rm sub}^{\Delta_{12}^m} &=& \frac{M_{\rm sub}^{\Delta_{12}^m}}{M_{200}} = 0.11 \\
\forall \frac{m_{200}}{M_{200}} &\in& 
\left[\frac{m_1}{M_{200}}=2.2\times 10^{-6},\frac{m_2}{M_{200}}=8.8\times 10^{-4}\right]\;.\nn
\een
We note that this is fully consistent with the semi-analytic result obtained in 
Ref.~\cite{Zentner2003}, which sets this fraction to $\sim$10\% for subhalos in the mass range 
$10^{-5}<m/M<10^{-2}$, assuming that $dN/dm \overset{\sim}{\propto} m^{-2}$ (see also
Refs.~\cite{Diemand2007,Kuhlen2008,Kamionkowski2010,Pieri2011}). This estimate is valid
for galactic halos of masses $M_{200}\sim 10^{12}\,M_\odot$, and may actually evolve with the host
halo mass as it increases up to the cluster mass scale (see \eg~Refs.~\cite{Gao2012a,Xu2015}).

In practice, we will match the fraction $f_{\rm sub}^{\Delta_{12}^m}$ defined in \citeeq{eq:subfrac}
to the above $\tilde{f}_{\rm sub}^{\Delta_{12}^m}$ by replacing the tidal mass $m_t$ by $m_{200}$ in
\citeeq{eq:mmeantilde}. An important subtlety is that we will only integrate over the subhalo 
population which has not been disrupted by tidal interactions with the host dark halo
(so-called global tides in \citesec{sssec:global_tides}).
Indeed, these interactions are at play in VL2, so this normalization procedure must take them
into account. Note that the calculation of the phase-space normalization factor $K$ defined in
\citeeq{eq:subnorm} must also include these tidal cuts, which are
position-mass-concentration dependent. In practice, this is done by integrating the concentration
function from a minimal concentration, $c_{\rm min}(m,R)\geq 1$, which is set by the tidal
disruption model and depends on the subhalo mass and its position in the Galaxy
(see \citesec{sssec:disruption}).

We emphasize that since VL2 is a DM-only simulation, the above normalization can only be used
to calculate the total number of subhalos {\em before} plugging in tidal stripping from the
baryonic components. This is actually very fortunate because this really allows us to
{\em predict} the baryonic effects (at least those related to tidal stripping), instead of trying
to reproduce them. Indeed, we stress that the way baryons are implemented in simulations is still
highly debated in numerical cosmology (see \eg~Ref.~\cite{WeinbergEtAl2015}). We will deal with
baryonic tides only in a second, independent step.

To summarize the normalization procedure, we first fix the total number of subhalos
{\em before baryonic tides} by matching $f_{\rm sub}^{\Delta_{12}^m}$ defined in \citeeq{eq:subfrac}
to the constraint $\tilde{f}_{\rm sub}^{\Delta_{12}^m}$ given in \citeeq{eq:normsub}
(replacing the tidal mass $m_t$ with $m_{200}$ in the definition of $f_{\rm sub}^{\Delta_{12}^m}$).
In a second step, we plug in baryonic tides, which turn out to be dominated by disk-shocking
effects. It is easy to show that the final number of subhalos $N'$ will merely be given by
$N'=(K'/K)\times N \leq N$, where $K$ ($K'\leq K$) and $N$ are the phase-space normalization and the
total number of objects, respectively, before (after) including baryonic tides. This relies
on matching the global subhalo mass density in the outskirts of the Galaxy, where baryonic effects
can be neglected. Tidal effects will be discussed in detail in \citesec{ssec:tides}.

\subsection{Global and internal subhalo properties}
\label{ssec:sub}
In this section, we specify the global and internal properties. The latter are mostly featured
by the inner density profile $\rho$ of a subhalo and its specific concentration $c$.
For the density profile, we assume spherical symmetry and adopt the NFW shape given by 
\citeeq{eq:dmprofile}, with $(\alpha,\beta,\gamma)=(1,3,1)$ as our default configuration, unless 
specified otherwise. We define the concentration parameter $c$ as
\ben
\label{eq:c200}
c=c_{200}= \frac{r_{200}}{r_{-2}}\,,
\een
where $r_{-2}$ is the radius at which the logarithmic slope $d\ln(\rho)/d\ln(r)=-2$. In the
$\alpha\beta\gamma$ case, we have
\ben
\kappa\equiv \frac{r_{-2}}{r_s} = \left\{\frac{(\beta-2)}{(2-\gamma)}\right\}^{-\frac{1}{\alpha}}\;
\forall \; 0<\gamma<2\,,\frac{\beta-\gamma}{\alpha}>2\,,\nn\\
\label{eq:rm2}
\een
such that $\kappa=1$ and $r_{-2}=r_s$ for an NFW profile. The same is readily obtained for an
Einasto profile. The concentration parameter will play a significant role not only in ruling the
subhalo annihilation rate, but also in characterizing the resistance of subhalos to tidal stripping.

We now formulate the overall mass and the tidal mass:
\ben
m&=&m_{200}=m(r_{200})= 4\,\pi\,r_s^3 \int_0^{\kappa\,c} dx\,x^2\,\rho(x\,r_s)\nn\\
m_t &=& m(r_t) = 4\,\pi\,r_s^3 \int_0^{x_t} dx\,x^2\,\rho(x\,r_s)\,\zeta(x_t)\,,
\label{eq:msub}
\een
where the dimensionless parameter $x\equiv r/r_s$, and $x_t\equiv r_t/r_s$ ($r_t$ is the tidal 
radius). Function $\zeta(x_t)$ takes values 0 or 1 to account for the potential tidal disruption
of the subhalo. We will specify this function as well as our definition of the tidal radius in 
\citesec{ssec:tides}. Note that this definition of the tidal mass implicitly assumes
  that the inner structure subhalos are not affected by tidal effects. We will further comment
on this approximation in \citesec{ssec:tides}.

In the same vein, we also introduce the subhalo effective 
annihilation volume $\xi$:
\ben
\xi&\equiv &\xi_{200}=\xi(r_{200})= 4\,\pi\,r_s^3 \int_0^c dx\,x^2\,
\left\{\frac{\rho(x\,r_s)}{\rho_0} \right\}^2\nn\\
\xi_t&=&\xi(r_t)= 4\,\pi\,r_s^3 \int_0^{x_t} dx\,x^2\,
\left\{\frac{\rho(x\,r_s)}{\rho_0} \right\}^2\,\zeta(x_t)\,,
\label{eq:xi}
\een
which provides a measure of the WIMP annihilation rate in a subhalo.
It actually quantifies the volume a subhalo would have to supply its annihilation rate 
if it were a homogeneous sphere of reference density $\rho_0$. In practice, we will set
$\rho_0=\rho_\odot$, unless specified otherwise. This is a particularly convenient choice
in the context of indirect DM searches with antimatter cosmic rays 
\cite{Lavalle2007a,Lavalle2008c,Pieri2011}. It is similar to the definition of the $J(\psi)$
luminosity factor in the context of gamma-ray searches \cite{Bergstroem1998b}.

We now introduce key physical quantities to describe bounded systems, which we will use when
addressing the tidal effects in \citesec{ssec:tides}. We first define the gravitational binding
energy, {\em i.e.} the minimum energy to unbound the system, as
\ben
E_b(r_t) = 4\,\pi\, G_N \int_0^{r_t} dr \,r \, \rho(r)\,m(r)\,,
\label{eq:ebind}
\een
where $r_t$ is the subhalo tidal radius, $\rho(r)$ its mass density at radius $r$, and
$m(r)$ its mass inside $r$; the binding energy is defined as positive. Alternatively, we
also introduce the potential energy of a bounded system:
\ben
U_g(r_t) = 2\,\pi\,G_N \int_0^{r_t} dr \,r^2 \, \rho(r)\,\tilde{\phi}(r)\,,
\label{eq:ug}
\een
where we have used the gravitational potential
\ben
\label{eq:grav_pot}
\tilde{\phi}(r) &=& \tilde{\phi}(r,r_t) \equiv \phi(r) - \phi(r_t)\\
\text{with}\;\phi(r) &=& -G_N\int_r^\infty dr\,\frac{m(r)}{r^2}\nn\,,
\een
taking into account that subhalos have finite extensions set by their tidal radii $r_t$.
This potential takes an analytic expression for an NFW profile, easy to derive and available in any
relevant textbook. Both the binding energy and the (absolute value of the) potential energy
scale similarly with $r_t$ for NFW profiles, very roughly $\propto r_t^2$ when $r_t\ll r_s$, and
$\propto\ln(r_t)$ when $r_t\gg r_s$.

In the following sections, we will provide more details on the overall global phase space
characterizing our subhalo population model. We will discuss the concentration function in
\citesec{sssec:dPdc},
the mass function in \citesec{sssec:dPdm}, the spatial distribution in \citesec{sssec:dPdV},
and tidal effects and induced correlations in \citesec{ssec:tides}.

As an important point, we will assume that subhalos are independent of each other, which means
that each physical quantity (mass, annihilation volume, \etc) can be dealt with as a random
variable over the global phase space. This will allow us to compute different moments of any
observable and thereby estimate the associated statistical uncertainty.
\subsubsection{Concentration function}
\label{sssec:dPdc}
The concentration of DM (sub)halos has long been studied in the literature 
(see \eg~Refs.~\cite{Bullock2001a,Eke2001,Wechsler2002,Maccio2007,Maccio2008,Springel2008,Prada2012,Sanchez-Conde2014,Dutton2014,Ludlow2014,Zhu2016}). In Ref.~\cite{Sanchez-Conde2014} (SCP14 hereafter), 
the authors compared the concentration model of Ref.~\cite{Prada2012} to various sets of 
cosmological simulation data, spanning a large range of subhalo masses, notably from 
$\sim 10^{-6}\msun$ (from Refs.~\cite{Diemand2005a,Anderhalden2013,Ishiyama2013}), and also 
including the VL2 data.
It turns out that in spite of the slightly different input cosmological 
parameters, these data can be relatively well described by the model within statistical errors ---
note that the rather large values of $\Omega_m$ and $\sigma_8$ inferred from the recent Planck data
would even favor a more optimistic modeling \cite{Dutton2014,Dooley2014}. The authors of SCP14 
also provide a fitting function of the central concentration value, inspired by
Ref.~\cite{Lavalle2008c}, which is quite convenient for our purposes:
\ben
\label{eq:cbar}
\bar{c}(m,z=0) = \sum_{i=0}^5 c_i\,\left[ \ln\left( \frac{m}{h^{-1}\,\msun}\right) \right]^i\,,
\een
with $c_i=[37.5153, −1.5093, 1.636 \times 10^{-2}, 3.66 \times 10^{-4},−2.89237 \times 10^{-5}, 
5.32 \times 10^{-7}]$, which gives values from $\sim 65$ at the lower subhalo mass edge 
$\sim 10^{-10}\msun$ to $\sim 10$ at the bigger mass edge $\sim 10^{10}\msun$. This is reminiscent
of the fact that smaller objects have formed earlier, in a denser universe, and this further 
induces a larger luminosity-to-mass ratio $\xi/m\overset{\sim}{\propto}c^3$ for lighter objects.

Furthermore, there is a scatter about this central value related to the fact that structure
formation is a statistical theory of initial density perturbations. The associated PDF can be very
well described by a log-normal distribution 
(see \eg~Refs.~\cite{Jing2000,Bullock2001,Wechsler2002,Maccio2007,Maccio2008}):
\ben
\label{eq:dpdc}
\frac{d{\cal P}_c}{dc}(c,m) = \frac{1}{K_c}
\frac{\exp\left\{-\frac{(c-\bar{c}(m))^2}{2\,\sigma_c^2} \right\}}{c\,\sqrt{2\,\pi\,\sigma_c^2 }}\,,
\een
where we will fix the variance in log space to $\sigma_c = \sigma_c^{\rm dec}\times\ln(10)$,
with $\sigma_c^{\rm dec}=0.14$, a mass-independent and rather generic value consistent with several 
detailed studies (\eg~Refs.~\cite{Bullock2001,Wechsler2002,Maccio2007}). Parameter $K_c=K_c(m)$
allows a normalization to unity over the range considered in this work, that we set in practice to 
$c\in[1,\exp(\ln(\bar{c}(m)) + 8\,\sigma_c)]$. The lower value $c_{\rm min}=1$ is constrained by 
the definition of $r_{-2}$, which is no longer consistent when the halo extent is found to be
smaller in the case of both NFW and Einasto profiles. This does not mean that subhalos for which one
cannot specify $r_{-2}$ are nonphysical, this is just a limit of our definition of the concentration
itself \cite{KhairulAlam2001,Diemand2007a}. However, this has no impact on the observables we will 
be dealing with in this article, for which only large values of the concentration will be 
relevant.

Note that, according to \citeeqss{eq:cbar}{eq:dpdc}, the central concentration 
$\bar{c}$ and the averaged concentration $\langle c \rangle$ do not coincide:
\ben
\label{eq:cmean}
\langle c(m) \rangle &=& \int dc \,c\,\frac{d{\cal P}_c}{dc}(c,m)\\ 
&\simeq& \bar{c}(m)\,e^{\frac{\sigma_c^2}{2}} \simeq 1.05\, \bar{c}(m) \neq \bar{c}(m)\,.\nn
\een

To summarize, once the density profile is fixed, the inner structure of a subhalo is fully 
determined from its mass $m$ and its concentration $c$. The former gives $r_{200}$ from
\citeeq{eq:m200}, and the latter provides the scale radius $r_s$ and the scale density 
$\rho_s$ from \citeeqss{eq:c200}{eq:msub}.

We emphasize that the concentration function introduced above has to be understood as a
{\em cosmological} function only valid to describe field subhalos, \ie\ subhalos which have not
been subject to tidal stripping yet and have retained information about their cosmological
origin. This function will actually be modified by tidal effects as we will see later. Indeed,
concentration will play a crucial role in characterizing the resistance of subhalos to tidal
effects. In our approach, tides will actually not modify the shape of the concentration function
defined in \citeeq{eq:dpdc}, but will erode the concentration range from the left (the less
concentrated objects will be disrupted more efficiently): the minimal concentration $c_{\rm min}$
will therefore become spatial-dependent and will strongly increase toward the central parts of
the Galaxy, such that the available phase-space volume will be strongly suppressed
(see \citesec{ssec:tides}).
\subsubsection{Mass function}
\label{sssec:dPdm}
An important part of the subhalo phase space consists in the mass function. The Press and Schechter
(PS) formalism and its extensions (see Refs.~\cite{Press1974,Bond1991,Lacey1993,Sheth2001,Cooray2002,Zentner2003,Zentner2007,Giocoli2008}), in the frame of hierarchical structure formation and 
standard cosmology, provide the basic theoretical paradigm to understand why cosmological 
simulations exhibit power-law (sub)halo mass functions down to very small subhalo masses 
(see \eg~Refs.~\cite{Diemand2006,Diemand2007,Diemand2008b,Springel2008,Zhu2016}). The mass index is 
actually related to the index of the power spectrum of primordial perturbations, and remains
weakly constrained on the very small scales relevant to DM subhalos (for recent studies, see
\eg~Refs.~\cite{Bringmann2012b,ClarkEtAl2016a}). However, we will still assume that the mass
function is regular over the whole subhalo mass range, as expected in standard cosmology, such
that the initial mass PDF may be written as a simple power law,
\ben
\frac{d{\cal P}_m}{dm} = K_m \,\left\{ \frac{m}{\msun} \right\}^{-\alpha_M} \;{\rm and}\;
\int_{m_{\rm min}}^{m_{\rm max}}dm\,\frac{d{\cal P}_m}{dm} = 1\,,\nn\\
\label{eq:dpdm}
\een
where $K_m=K_m(m_{\rm min},m_{\rm max})$ allows the normalization of the PDF to unity over the
mass range delineated by $[m_{\rm min},m_{\rm max}]$. Note that we implicitly assume $m=m_{200}$. 
The mass index $\alpha_M$ is typically expected to be $\lesssim 2$ as a prediction of the PS theory 
with standard cosmological parameters, which is actually recovered in cosmological simulations 
\cite{Diemand2006,Diemand2007,Diemand2008b,Springel2008,Zhu2016}. In the following, we will
assume $1.9\leq \alpha_M\leq 2$, unless specified otherwise.

We emphasize that the {\em actual} subhalo mass distribution, which should incorporate tidal 
stripping and disruption, and depends on the tidal subhalo mass $m_t$ rather than on $m_{200}$, is 
not {\em directly} described by \citeeq{eq:dpdm}. Indeed, tidal effects will make $m_t$ become
position dependent, and thereby the subhalo mass range too. Nevertheless, the procedure presented
in \citesec{ssec:halo_and_sub} (see \citesec{sssec:overall_sub} and \citesec{sssec:norm_sub}) 
includes all of this self-consistently while still being based on \citeeq{eq:dpdm} as the
{\em initial} mass function.

Finally, we remind the reader that the minimal, or cut-off mass $m_{\rm min}$ (that may also
further appear as $m_{\rm cut}$) is linked to mean free path of DM particles per Hubble time in the
early universe at the time of matter-radiation equivalence, which fixes the minimal size of
the structures which can grow under gravity \cite{Peebles1984,Silk1993,Kolb1994a,Chen2001,Hofmann2001,Berezinsky2003,Green2004a,Loeb2005,Boehm2005a,Profumo2006,Bertschinger2006a,Bringmann2007,Bringmann2009,Berezinsky2014,Visinelli2015}. This mass scale is therefore related to the scattering properties of DM particles, and for WIMPs, typical values are $\sim 10^{-6}\,M_\odot$ for 100 GeV particle
masses, down to $\sim 10^{-10}\,M_\odot$ for TeV particle masses \cite{Bringmann2009}.

\subsubsection{Spatial distribution}
\label{sssec:dPdV}
The spatial distribution of subhalos in the Galaxy is a very important input in this work
because it will allow us to compute the local number density of subhalos, which will itself
set the local annihilation boost factor, relevant, for instance, to indirect DM searches
with antimatter cosmic rays. As for the mass function introduced above, we will also define the
{\em initial} spatial distribution, which will further be distorted by tidal effects from the
procedure defined in \citesec{ssec:halo_and_sub}. As argued above, since small-scale subhalos have
already virialized when the Galaxy forms, it is reasonable to match the initial spatial PDF to the
global dark halo profile, such that
\ben
\label{eq:dpdv}
\frac{d{\cal P}_V(\vec{x})}{dV} = \frac{\rho_{\rm tot}(\vec{x})}{M_{200}}\,,
\een
where $M_{200}$ is the global dark halo mass within $R_{200}$, and $\rho_{\rm tot}$ is the global DM
density profile discussed in \citesec{sssec:darkhalo}. This PDF is normalized to unity within a
sphere or radius $R_{200}$ by construction.

Of course, tidal effects will strongly distort this initial distribution because of tidal
disruption, such that the effective and {\em real} spatial distribution of subhalos
will eventually not look like \citeeq{eq:dpdv}. Actually, tidal effects will make this spatial
distribution become mass dependent, exactly as the {\em actual} mass function becomes spatial
dependent, such that the mass and spatial distributions are fully intricate in practice
(tidal effects are discussed in \citesec{ssec:tides}). Therefore,
even though we do use \citeeq{eq:dpdv} to formally describe the initial spatial distribution, the
effective spatial distribution is still self-consistently determined through the procedure
described in \citesec{sssec:overall_sub} and \ref{sssec:norm_sub}.

\subsection{Tidal effects}
\label{ssec:tides}
Tidal effects play a fundamental role in shaping the phase space relevant to Galactic DM subhalos as
defined in \citeeq{eq:subps}. As discussed above, they affect their mass, concentration, and spatial
distributions, and will thereby distort and mix the PDFs defined in \citeeqss{eq:dpdm}{eq:dpdc} by
pruning and disrupting subhalos. In the following, we describe in detail the way we implement these
effects, which are critical to our final results.

Many studies have been, and are still being, carried out on this topic 
(\eg~Refs.~\cite{Tormen1998,Taylor2001,Bullock2001b,Hayashi2003,Zentner2005,Goerdt2007,Berezinsky2008,Kazantzidis2009,DOnghia2010,Gan2010,Bartels2015,Emberson2015,Jiang2016,vandenBosch2016,Han2016b,Zhu2016,Moline2016}).
In this study, we will mostly consider two distinct effects: tidal stripping from the overall 
Galactic potential, and tidal shocking by the Galactic disk, which are known to be the most
significant processes (see \eg~Ref.~\cite{Berezinsky2014}).

Implicit in what follows, we will assume that any derived tidal radius cannot exceed $r_{200}$,
such that formally, throughout all this paper, we will always impose
\ben
r_t \leq r_{200}\,.
\een

We will further consider circular orbits, and assume that the internal structure of a
subhalo is not affected inside $r_t$, which is consistent with the circular orbit approximation.
Actually, a very simple reasoning is enough to convince oneself that tidal effects can also
remove particles from the inner regions of a subhalo. For instance, a gravitationally bound
spherical system with maximal symmetry (\eg\ an ergodic system) has a central phase space that
can in principle explore velocities up to the escape speed. So even if this concerns the very
tail of the particle distribution, a small acceleration applied to this high-speed population is
enough to remove particles from the center. This would even be more efficient in systems with a
large fraction of eccentric orbits. However, even though some particles from regions within $r_t$
should indeed be kicked out from subhalos because of tidal stripping, we expect their fraction
to be a second-order correction to our results, because these particles are located on the
phase-space tails. This approximation is expected to be more and more reliable as the
concentration of objects increases, \ie\ as their initial cosmological mass decreases, which
is the mass range of interest in boost factor calculations. This is actually confirmed by several
dedicated simulation studies (see \eg~Refs.~\cite{Hayashi2003,Emberson2015}).
\subsubsection{Global tides from the host halo}
\label{sssec:global_tides}
Tidal effects generated by the host Galactic halo induce a pruning of subhalos that can be
accounted for by setting the actual spatial extent of a subhalo to its tidal radius. In the
simplest approximation where both the host halo and its subhalo are considered as pointlike
objects, and taking into account the centrifugal force, the tidal radius can be defined as 
\cite{King1962a,Binney2008,Hayashi2003,Read2006a}
\ben
r_{t\bullet} &=& r_{t\bullet}(R,m,M) = \left\{ \frac{m_t}{3\,M}  \right\}^{1/3} \,R\,,
\label{eq:rt_jacobi}
\een
where $M$ and $m_t$ are the point masses of the whole host galaxy and the subhalo, respectively,
and $R$ is the radial position of the subhalo in the host galaxy. Note that $m_t= m(r_{t\bullet})$
is featured in the above equation, not $m_{200}$.
We will refer to the above definition of the tidal radius as the {\em pointlike Jacobi
  limit}.
This formula can be generalized to the case of objects 
orbiting galaxies with continuous mass density profiles, more relevant to our case, as 
(see Ref.~\cite{Binney2008})
\ben
\label{eq:rt}
r_t &=& r_t(R,m_t,\rho_{\rm tot}(R)) \\
&=& 
\left\{ \frac{m_t}{ 3\,M(R)\, \left(1-\frac{1}{3}\frac{d\ln M(R)}{d\ln(R)}\right)}  \right\}^{1/3} \,R\,,\nn
\een
where $M(R)$ is the host galaxy mass within a radius $R$, which depends on the global mass density 
profile $\rho_{\rm tot}$. This equation may be solved iteratively as it implies the tidal subhalo 
mass $m_t=m(r_t)$ defined in \citeeq{eq:msub}, and is shown to provide a rather good description of
a subhalo radial extent in DM-only cosmological simulations (see \eg~Ref.~\cite{Springel2008}).
We will refer to this definition of the tidal radius as the {\em smooth Jacobi limit}.

For completeness, we may also introduce an empirical tidal radius definition where we just
delineate the subhalo by the radius at which its density equals the overall density locally,
{\em i.e.}
\ben
r_t\;\text{such that}\;\rho(r_t)=\rho_{\rm tot}(R)\,.
\label{eq:rt_dens}
\een
We will refer to this definition of the tidal radius as the {\em isodensity tidal radius}.

Finally, we stress that when baryons are included, they also contribute to $\rho_{\rm tot}$ and
thereby to $M(R)$ in the equations above [for the baryonic disk, we will use the spherical
approximation of the density, given in \citeeq{eq:sph_disk}]. We will discuss the impact of
using one or the other definition in \citesec{sssec:disruption}. Besides this, note that although
global tides from the host halo are indeed important in the outskirts of the Galaxy, other
processes become more and more efficient in the inner regions, as the ratio of baryons to dark
matter increases, as we will see below.

\subsubsection{Baryonic disk shocking}
\label{sssec:disk_tides}
An important source of destructive gravitational interaction arises during disk crossing, where
subhalos can acquire a substantial amount of kinetic energy which can unbind them
(see Refs.~\cite{Ostriker1972,Gnedin1999b,Taylor2001,Berezinsky2008,DOnghia2010}). Termed
{\em disk shocking}, this effect dominates over more local destructive effects like encounters
with stars, and is actually the most efficient subhalo disruption mechanism in the luminous
part of spiral galaxies \cite{Berezinsky2014}. These effects are much more tricky to include than
those discussed in \citesec{sssec:global_tides}.

Below, we discuss the physical steps that allow us to account for disk shocking in a subhalo
population model. We first review the seminal results obtained in Ref.~\cite{Ostriker1972} by
Ostriker, Spitzer, and Chevalier, and further extended in \eg~Ref.~\cite{Gnedin1999b}, which were
related to the study of Galactic stellar clusters.

We wish to evaluate the kinetic energy gained by a WIMP orbiting a subhalo only subject to the
gravitational field of the Galactic disk during one crossing. Assuming the disk is an infinite
slab (radial boundaries are sent to infinity), then the
disk gravitational force field is directed along the axis perpendicular to the disk and sustained
by the unitary vector $\vec{e}_z$, so the $z$ coordinate is the only relevant one here. This is a
fair approximation when a subhalo is about to cross the disk. Setting $\vec{x}$ as the full 3D
WIMP position and $\vec{x}_0$ as the subhalo center position, the change in the WIMP velocity
along the $z$ axis and in the subhalo frame reads
\ben
\frac{dv_z}{dt} &=& \frac{d(\dot{\vec{x}} - \dot{\vec{x}}_0)}{dt}\cdot \vec{e}_z
= g_{z,{\rm disk}}(Z)- g_{z,{\rm disk}}(Z_0)  \nn\\
&=& g_{z,{\rm disk}}(Z_0+\delta Z)-g_{z,{\rm disk}}(Z_0)\nn\\
&\approx& \delta Z \,\frac{d g_{z,{\rm disk}}(z)}{dz}\,,
\label{eq:dvdt}
\een
where we have defined $\delta Z\equiv Z-Z_0 = (\vec{x} - \vec{x}_0)\cdot \vec{e}_z$, and
where the latest line is obtained from a simple Taylor expansion to first order. We have
used the disk gravitational force field $g_{z,{\rm disk}}$, which can be inferred from the
baryonic disk profile introduced in \citeeq{eq:rho_baryons},
\ben
|g_{z,{\rm disk}}(R,z)| = 4\,\pi\,G_N\,z_d\,\rho_d(R,z)\;.
\label{eq:gravff}
\een

\citeeq{eq:dvdt} can further be integrated over the disk crossing time $\delta t=t_>-t_<$ to get
the net velocity change $\Delta v_z$,
\ben
\label{eq:speed_change}
\Delta v_z &=& \int_{t_<}^{t_>} dt\, \delta Z \, \frac{d g_{z,{\rm disk}}(z)}{dz}\\
&\approx& \delta t \, \delta Z \,
\frac{(g_{z,{\rm disk}}(z(t_>))-g_{z,{\rm disk}}(z(t_<))}{z(t_>)-z(t_<)}\nn\\
&=& \delta Z \, \left\{   \frac{\delta t}{z(t_>)-z(t_<)}=\frac{1}{V_z}\right\} \times 2\,g_{z,{\rm disk}}(z=0)\,,\nn
\een
where $V_z$ is the component of the subhalo velocity perpendicular to the disk. This approximation
is licit as long as $\delta Z$ does not vary much over the crossing time ({\em i.e.} the WIMP
orbital time in the subhalo is much longer than the disk crossing time) and as long as
the modulus of the gravitational force field remains close to its maximal value (aside from the
flip of sign when crossing $z=0$). This is known as the {\em impulsive approximation}.

We can therefore derive the net average gain in kinetic energy per unit WIMP mass for a single disk
crossing,
\ben
\epsilon^0_k(z) &\equiv& \frac{\Delta E_k^0}{m_\chi} = \frac{1}{2} (\Delta v_z)^2\nn\\
&=& \frac{2\,g^2_{z,{\rm disk}}(z=0)\,z^2 }{V_z^2}\,,
\label{eq:diffkin_noA}
\een
which depends on the squared vertical coordinate $z$ relative to the subhalo center.

A key assumption in deriving the previous results is that $\delta Z$ does not vary significantly
as the subhalo crosses the disk. This is very likely not verified for the innermost orbits, nor
for the smallest objects, for which the impulsive approximation readily breaks down. Indeed, had
subhalo particles enough time to circulate several times about the center as the object crosses
the disk, conservation of angular momentum would prevent them from leaving the system, and disk
shocking would become inefficient. This is an example of the manifestation of adiabatic invariance,
which was extensively studied in the context of stellar clusters in
Refs.~\cite{Weinberg1994,Weinberg1994a,Gnedin1999,Gnedin1999b,Gnedin1999a}, from both analytic
and numerical calculations. Following Ref.~\cite{Gnedin1999b}, capturing the results derived in
Ref.~\cite{Weinberg1994a} from the linear theory approximation, we introduce an adiabatic
correction,
\ben
A(\eta) = (1+\eta^2)^{-3/2}\,,
\label{eq:adiab}
\een
where $\eta$ is the so-called {\em adiabatic parameter}, with $\eta\gg 1$ for orbits close to the
object's center, and $\eta\ll 1$ close to the tidal radius. This gives $A(\eta\gg 1)\to 0$, and
$A(\eta\ll 1)\to 1$,
the latter case corresponding to the the parameter space for which the impulsive approximation
holds. The adiabatic parameter is formally defined as
\ben
\eta(r,R) \equiv \omega(r) \,\tau(R)\,,
\label{eq:eta}
\een
where $\omega$ is the orbital frequency that can be estimated from the inner dispersion velocity,
$\omega = \sqrt{\langle v^2\rangle(r)}/r$, with $r$ being the distance to the subhalo center, and
$\tau$ being the effective crossing time. The latter is given in terms of the half-height $H$ of
the disk, and of the vertical component of the subhalo velocity $V_z(R)$ at radius $R$ in the
Galactic frame. In the following, we will make use of the isothermal approximation, such that
each Cartesian component of the velocity dispersion, for any system of mass $m(r)$ inside a
radius $r$, is related to the circular velocity according to
\ben
\sigma_{v,i}^2(r) = \frac{1}{2}\,v_c^2(r) = \frac{1}{2}\,\frac{G_N\,m(r)}{r}\,.
\label{eq:isoth_approx}
\een
Consequently, we get
\ben
\label{eq:adiab_omega}
\omega(r) &\equiv& \sqrt{\frac{3\,G_N\,m(r)}{2\,r^3}}\\
&\approx& 
9.7\times10^{-2}{\rm Myr}^{-1} \sqrt{\frac{m(r)/\left\{m(r_s)=6\times 10^{-8}M_\odot\right\}}
    {\left(r/\left\{r_s=3.5\times 10^{-3}\,{\rm pc}\right\}\right)^3}}  \nn
\een
and
\ben
\label{eq:adiab_tau}
\tau(R) &\equiv & \frac{H}{V_z(R)} = H\,\sqrt{ \frac{2\,R}{G_N\, M(R)} }\\
&\approx& 0.45\,{\rm Myr}\,\frac{(H/100\,{\rm pc})}{(V_z/200\,{\rm km/s})}\,.\nn
\een
In \citeeq{eq:adiab_omega}, $m(r)$ stands for the subhalo mass inside a radius $r$, while
$M(R)$ featuring \citeeq{eq:adiab_tau} is the total Galactic mass inside a radius $R$.
We evaluated the orbital frequency $\omega$ for the mass a template subhalo of
$m_{200}=10^{-6}M_\odot$ has inside its scale radius $r_s\approx 3.5\times 10^{-3}$ pc ($m(r_s)
\approx 6\times 10^{-8}M_\odot$), taking the corresponding median concentration from
\citeeq{eq:cbar} ($\bar c\approx 60$). This shows that except in the very central parts of
subhalos where $A(\eta)\to 0$, we will essentially have $A(\eta)\sim 1$, corresponding to a
maximal efficiency for disk shocking. Nevertheless, since $m(r_s)/r_s^3\propto c^3$, we see that
this efficiency will decrease as the concentration increases, protecting the most concentrated
objects from disk-shocking effects. Actually, for a flat Galactic velocity curve of $\sim 200$
km/s, we find assuming an NFW profile that to get $\eta>1$, condition for the disk-shocking
efficiency to start to be damped out, one needs $x=r/r_s\lesssim 10^{-3}c$, regardless of the
subhalo mass.

The adiabatic correction $A(\eta)$ allows to modify the kinetic energy transfer defined in
\citeeq{eq:diffkin0} in such a way that it is now valid over the full extent of any considered
subhalo. This reads
\ben
\epsilon_k(z) &\equiv&  \frac{2\,g^2_{z,{\rm disk}}(z=0)\,z^2 }{V_z^2}\,A(\eta)\,,
\label{eq:diffkin0}
\een
where the vertical subhalo velocity component $V_z(R)$ has been implicitly defined in
\citeeq{eq:adiab_tau}.

Finally, assuming circular orbits for WIMPs in a subhalo, one can easily express the average
kinetic energy gain as a function of the radius $r$ only, as
$\langle z^2\rangle = (1/2)\int d\cos\theta\,r^2\,\cos^2\theta = r^2/3$. We get
\ben
\langle \epsilon_k\rangle (r) = \frac{2\,g^2_{z,{\rm disk}}(z=0)\,r^2 }{3\,V_z^2(R)}\,A(\eta)\,.
\label{eq:diffkin}
\een
The scaling with $r$ is explicit, except for the quasi-exponential suppression when $r\to 0$ due
to adiabatic invariance: the gain in kinetic energy increases like the squared radius, and
is maximal close to the tidal boundary of the subhalo. This scaling is shown as red curves in
\citefig{fig:diffds} for two different subhalo masses, $10^{-6}$ (solid curve) and $10^{6}M_\odot$
(dashed curve), and further compared to the moduli of their gravitational potentials, defined in
\citeeq{eq:grav_pot}.

\begin{figure}[!t]
\centering
\includegraphics[width = 0.495\textwidth]{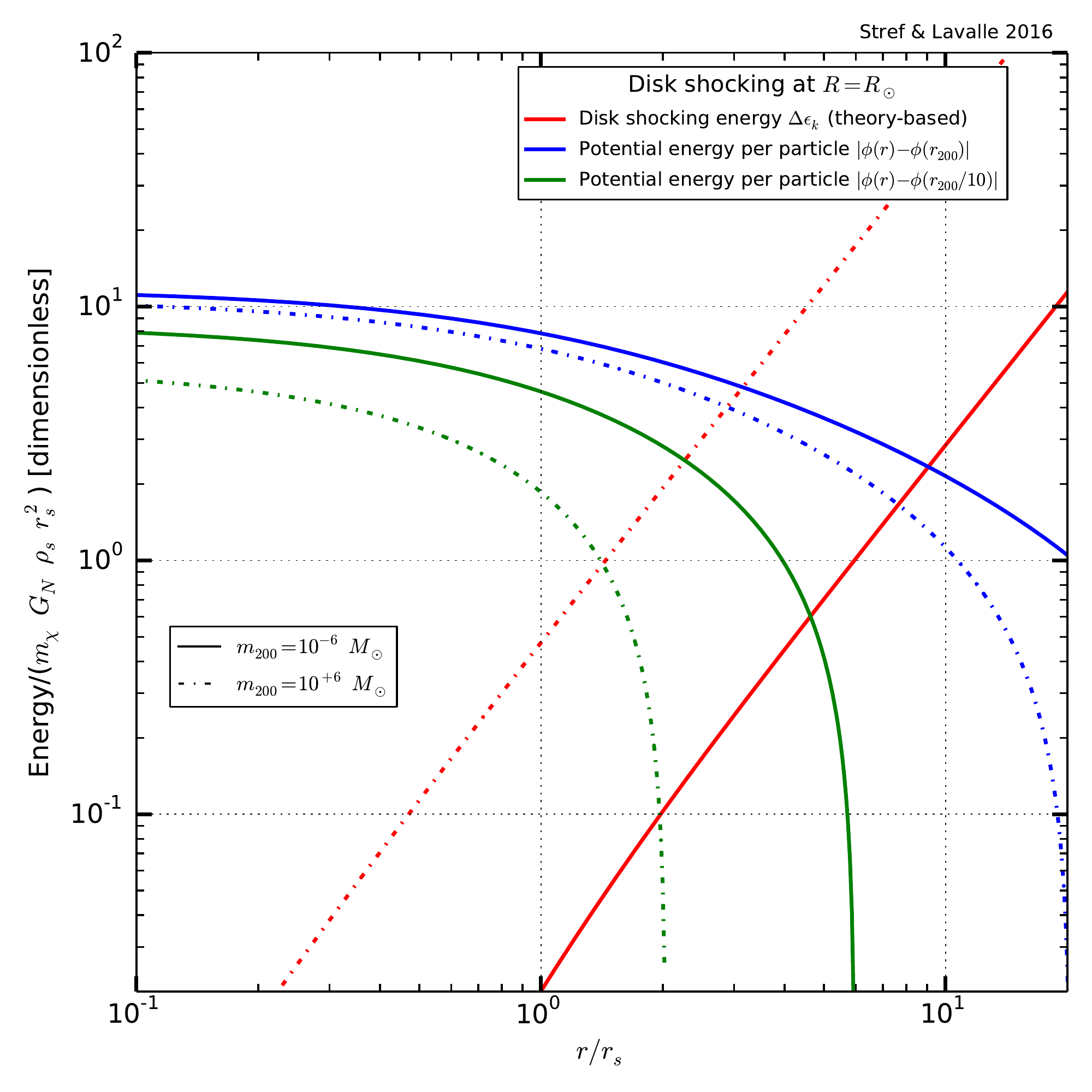}
\caption{\small Average kinetic energy gain induced by a single encounter with the Galactic disk as
  a function of $r/r_s$, for two NFW subhalos ($10^{-6}$ and $10^{6}M_\odot$) located at $R_\odot$ --
  see the definition in \citeeq{eq:diffkin} (here, we use the M11 mass model parameters). Also
  shown are their gravitational potentials. We plot the results in units of $G_N\,\rho_s\,r_s^2$,
  which comes out as a natural scaling, to make the comparison more striking.}
\label{fig:diffds}
\end{figure}

The calculations presented above are at the basis of the methods we propose to follow to
account for disk shocking, and thereby to further prune or destroy subhalos. Below,
we discuss two different strategies, which we will call {\em differential} and {\em integrated}
disk shocking to make the distinction clear. Common to both methods is the number of disk crossings,
$N_{\rm cross}$, which is computed from the circular velocity of a subhalo in the Galactic frame
(we implicitly assume circular orbits) and the age of the Galaxy $T_{\rm MW}$:
\ben
N_{\rm cross}(R) = \sqrt{\frac{G_N\,M(R)}{R}}\, \frac{T_{\rm MW}}{\pi\,R}\,.
\label{eq:ncross}
\een
Throughout this paper, we will set $T_{\rm MW}=10$ Gyr. Assuming the M11 Milky Way mass model
described in \citesec{ssec:dynamics}, the number of disk crossings is $\sim(670,92,37)$ for
Galactocentric radii $R\in(1\,{\rm kpc},R_\odot,20\,{\rm kpc})$, respectively. This already
tells us that disk shocking will lead to more efficient tidal stripping for subhalos which
venture toward the central parts of the Galaxy.

Besides the disk-shocking methods presented below, which are aimed to determine subhalo tidal
radii in Galactic regions encompassing the baryonic disk, our tidal {\em disruption} criteria will
be discussed in \citesec{sssec:disruption} where the disk-shocking methods will be further compared.

\paragraph{Tidal radius from differential disk shocking.}~\\

The so-called differential disk-shocking method will be our primary method, and relies on a
comparison between the kick in velocity induced by disk shocking, as effectively described in
\citeeq{eq:diffkin}, and the escape velocity
\ben
v_{\rm esc}(r) = \sqrt{-2\,\tilde{\phi}(r)}\,.
\label{eq:vesc}
\een
If the kick induced by disk shocking is such that the particle reaches the escape velocity, then
it gets unbound to the system. Therefore, for each disk crossing, we will accordingly define the
tidal radius as the radius at which the kick in velocity equals the escape velocity. In terms of
energies, this reads
\ben
r_t\;\text{such that}\;\langle \epsilon_k\rangle (r_t) =  - \tilde{\phi}(r_t)\,.
\label{eq:rt_diffds}
\een
This procedure must be applied at each crossing, such that it may somehow capture the dynamics
of disk shocking. Indeed, hidden in $\tilde{\phi}$ [see \citeeq{eq:grav_pot}] is the radial
boundary of the subhalo, which means that the above equation must be applied iteratively up to
the number $N_{\rm cross}$ of disk crossings given in \citeeq{eq:ncross}. More explicitly,
we have for the $i^{\rm th}$ crossing
\ben
r_{t,i}\;\text{such that}\;\langle \epsilon_k\rangle (r_{t,i}) =  - \tilde{\phi}(r_{t,i},r_{t,i-1})\,.
\label{eq:rt_diffds_it}
\een
In practice, we start with the tidal radius inferred from the global tidal effects induced
by the host halo and discussed in \citesec{sssec:global_tides}.
This method can easily be applied to any subhalo model, irrespective of the inner density profile.
It also provides a dynamical description of disk shocking, while only approximately. Indeed,
this iterative procedure assumes that the internal structure of the shocked subhalo is not
altered between two crossings, while part of the energy could actually be redistributed. Anyway,
this picture is still consistent with adiabatic invariance, which partly protects the inner parts
of subhalos against tidal pruning.

An illustration of this differential disk-shocking method is shown in \citefig{fig:diffds},
where we have plotted the disk-shocking energy $\langle \epsilon_k\rangle (r)$ (red curves)
and the gravitational potential modulus $|\tilde{\phi}(r)|$ as a
function of the scaling variable $r/r_s$ (where $r_s$ is the subhalo scale radius). We have
considered two different NFW subhalos, $10^{-6}$ (solid curves) and
$10^{6}M_\odot$ (dashed curves), both located at $R_\odot$. The corresponding
gravitational potential moduli are evaluated using two different radial boundaries for subhalos, one
set to $r_{200}$ (blue curves), and the other set to $r_{200}/10$ (green curves), above which they are
exponentially suppressed --- the $1/r$ scaling expected beyond
$r_s$ is poorly seen as the potential goes from $\propto {\rm cst}$ to
$\propto(1/r-1/r_t)\approx e^{-r/r_t}/r$ very fast. These radial boundaries can be thought of
as initial tidal radii before disk crossing such that the blue curves illustrate the potential
energies before the first crossing, while the green curves show how they have evolved after one or
several crossings. By virtue of \citeeq{eq:rt_diffds}, the tidal radius after one disk crossing
will be set to the radius at which the kinetic energy and the potential curves intersect.
Therefore, \citefig{fig:diffds} nicely illustrates why the tidal stripping efficiency is much
larger (i) in the outer regions of the system (compare where the red curves intersect the blue
curves -- first crossing), and (ii) for more massive subhalos (compare the relative
level at which the red curves intercept the green curves -- subsequent crossings).
The former effect is a consequence of the the regular increase of the differential disk-shocking
energy as $\propto r^2$ which will at some point encounter the decreasing potential, stripping
off the right-hand part of the DM content (the shift between the $10^{-6}/10^{6}\,M_\odot$
red solid/dashed curves is merely due to the difference in $r_s$); instead, the latter effect is
due to the fact that lighter subhalos are either much less extended and more concentrated, such
that after the first crossings, the further reduced disk shocking energy (because of the smaller
internal radius $r$, as it scales $\propto r^2$) makes only a tiny fraction of the residual
potential and only prunes the very external parts of small subhalos.

All this shows, in particular, that the impact of the number of crossings is important, though
quite not linear in this differential approach. The implementation of this method will represent
our primary tool to account for disk-shocking effects.

\paragraph{Tidal radius from integrated disk shocking.}~\\
\begin{figure}[!t]
\centering
\includegraphics[width = 0.495\textwidth]{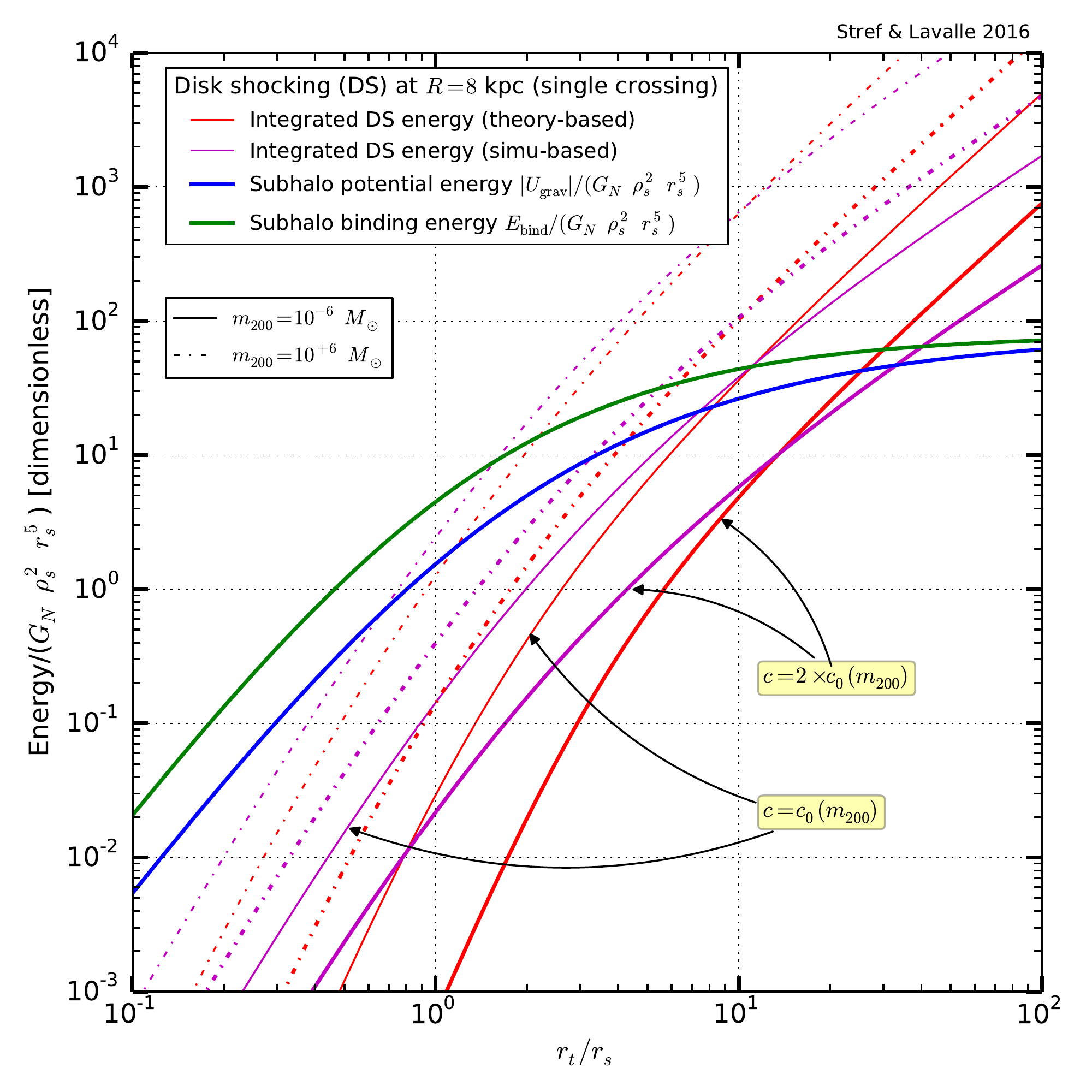}
\caption{\small Integrated kinetic energy gain induced by the Galactic disk as
  a function of $r_t/r_s$, for two NFW subhalos ($10^{-6}$ and $10^{6}M_\odot$) located at 8 kpc,
  assuming two different concentration values (the median one, and twice the median)
  --- see the definitions in
  \citeeq{eq:intds} for the theory-based calculation, and see \citeeq{eq:intds_simu} for the
  simulation-based estimate. Also shown are the binding and potential energies. We plot the results
  in units of $G_N\,\rho_s^2\,r_s^5$, which leads to the same binding and potential energies
  for both subhalo masses.}
\label{fig:intds}
\end{figure}
In contrast to what was presented above as a {\em differential} disk-shocking method, we can
now try to integrate the kinetic energy gain over the whole subhalo -- hence the term
{\em integrated} disk shocking method. Such a method was partly
followed in Ref.~\cite{Berezinsky2008}, where the authors used the Eddington equation
in the isothermal limit to convert the energy gain in phase space into a mass loss.
Here, instead, we will use spherical symmetry, and simply assume that WIMPs take only circular
orbits, such that the integrated kinetic energy gain can be expressed as
\ben
E_k(r_t,R) = 2\,\pi\int_0^{r_t} dr\,r^2 \int_{-1}^{1}d\cos(\theta)\,
\epsilon_k(z,R) \,\frac{\rho(r)}{m_\chi}\,,\nn\\
\label{eq:intds}
\een
where $\epsilon_k(z,R)$ is given by \citeeq{eq:diffkin0}, and $\rho$ is the inner subhalo
mass density profile. Spherical symmetry merely implies that $z^2=r^2\cos^2(\theta)$, which
makes the computation easy. This integrated energy gain can then be compared to the
binding energy or to the potential energy, for each subhalo. For an NFW profile, the
scaling goes from roughly $\propto \rho_s \, r_s^5 \,(r_t/r_s)^4$ for $r_t\ll r_s$, to
$\propto \rho_s \, r_s^5 \,(r_t/r_s)^2$ for $r_t\gg r_s$.

This is illustrated for a
single disk crossing in \citefig{fig:intds}, for two subhalos of $10^{-6}$ (solid curves)
and $10^{6}M_\odot$ (dash-dotted curves) located about the solar position -- for completeness,
we use two different concentration values for each subhalo: the median value (thin curves),
and twice the median (thick curves). The
red curves show the integrated disk-shocking energy given in \citeeq{eq:intds}, as compared
to the binding (green curve) and potential (blue curve) energies  [see \citeeqss{eq:ebind}{eq:ug},
  and the associated comments about the radial scaling]. Using units of
$G_N\,\rho_s^2\,r_s^5$ allows us to get a single curve for each of the latter energies,
for both subhalo masses.
Again, the increase in the disk-shocking energy is such that it will encounter the
binding energy, which flattens beyond $r_s$, thereby setting a tidal radius. As well depicted from
\citefig{fig:intds}, smaller as well as more concentrated objects will be less prone to tidal
stripping, which can be understood from working out the scaling $E_b(r_t)/E_k(r_t)\propto
\rho_s\,(r_t/r_s)^{-a} \propto c^3\,(r_t/r_s)^{-a}$, where $a\approx 3$ [see discussion just below
  \citeeq{eq:intds}]. This explains why, like it was already the case for differential disk
shocking, more massive subhalos will be more efficiently affected by tidal stripping.

Since we are dealing with integrated energies, we define the subhalo tidal radius after
$N_{\rm cross}$ disk crossings as
\ben
r_t\;\text{such that}\;N_{\rm cross} \,E_k (r_t,R) =  E_b(r_t)\,,
\label{eq:rt_intds}
\een
where we use the binding energy $E_b$ defined in \citeeq{eq:ebind} as a reference.

\paragraph{Tidal radius from integrated disk shocking (fits on cosmological simulations).}~\\
For the sake of comparison, we now introduce a result fitted on dark-matter-only zoomed-in
cosmological simulations, given in Ref.~\cite{DOnghia2010}, wherein a baryonic disk potential
was grown adiabatically to study the induced tidal disruption of subhalos. The qualitative features
of this result were recently recovered in cosmological simulations including baryons, and
discussed in Ref.~\cite{Zhu2016}. The authors of Ref.~\cite{DOnghia2010} have tried to capture
disk-shocking effects by a simple and physically motivated ansatz, which, as they found, matches
rather well with their simulation results (see \eg~Ref.~\cite{Binney2008} for the dynamical
grounds). They introduced an integrated-like disk-shocking energy $\tilde{E}_k(r_t,R)$ given by
\ben
\frac{\tilde{E}_k(r_t,R)}{E_b(r_t)} = \frac{(1.84\,r_{1/2})^2\,g_{z,{\rm disk}}^2}{3\,\tilde{\sigma}_v^2\,V_z^2}\,,
\label{eq:intds_simu}
\een
where $r_{1/2}$ is the radius containing half the subhalo mass, $g_{z,{\rm disk}}$ is the disk
gravitational force field given in \citeeq{eq:gravff}, $\tilde\sigma$ is an estimate
of the internal dispersion velocity given by $\tilde\sigma^2 = 0.4\,G_N\,m_t/r_{1/2}$, and
$V_z$ is the velocity component perpendicular to the disk, which will be inferred from the
approximation given in \citeeq{eq:isoth_approx}. This disk-shocking energy is shown as the
purple curves in \citefig{fig:intds}, for the two subhalo prototypes introduced above.
It still scales more sharply with $r_t$ (readily inferred as $\propto r_t^3$ from the equation
just above) than the potential or binding energy, though less sharply than the integrated
disk-shocking energy discussed in the previous paragraph, while still with a similar amplitude
around $r_t/r_s\simeq 10$. This means that this way to implement disk shocking will likely
disrupt subhalos more efficiently, as gravitational stripping toward the central regions
becomes more efficient. Still, we note that \citeeq{eq:intds_simu} relies on fits on simulation
results, and could therefore be more specific to the subhalo mass range probed by cosmological
simulations, which is still strongly limited by resolution issues. Anyway, the resulting subhalo
tidal radius after $N_{\rm cross}$ disk crossings can then be calculated by means of
\citeeq{eq:rt_intds}, by simply replacing $E_k(r_t,R)$ with $\tilde{E}_k(r_t,R)$.

\paragraph{Disk-shocking summary.}~\\
We have introduced the so-called {\em differential} and {\em integrated} disk-shocking energies.
For the latter, we have derived two expressions, one consistent with the differential one,
and another inspired by cosmological simulation and fully independent. These physical
quantities allow us to derive the subhalo tidal radius $r_t$ after $N_{\rm cross}$ disk crossings
for any method. These calculations lead to different results, but common to all is the fact
that $r_t$ does depend simultaneously on the subhalo mass $m$, its concentration $c$, its position
in the Galaxy $R$, and its internal density profile. Our primary method will be the one based on
the differential disk-shocking energy, as it relies on fewer assumptions. We will compare
all these results in \citesec{sssec:disruption}.

\subsubsection{Subhalo mass independence of $x_t=r_t/r_s$}
\label{sssec:xt}
A striking property of {\em all} the tidal radius calculation methods discussed above, both those
involving global tides and those involving disk shocking, is that the ratio $x_t=r_t/r_s$
turns out to be independent of the subhalo mass. Actually, $x_t$ depends only on the subhalo
concentration $c$ and on its radial position $R$ in the Galaxy. If the latter dependence is rather
easy to understand (tidal stripping depends on the position), the former is much less trivial.

For the global tides discussed in \citesec{sssec:global_tides}, it is easy to show that the methods
based on the Jacobi limit can be formulated along
\ben
x_t &=& \left[ \frac{\Delta_{200}\,f(x_t)}{\Delta_t(R)\,f(c)} \right]^{1/3}\,\kappa\,c\nn\\
\Leftrightarrow x_t\,\left[ f(x_t)\right]^{-1/3} &=&
\left[ \frac{\Delta_{200}}{\Delta_t(R)\,f(c)} \right]^{1/3}\,\kappa\,c\,,\nn
\een
which makes it clear that $x_t$ is only a function of $R$ and $c$. Here, $\kappa=r_{-2}/rs$ is
set by the choice of the inner profile, and the function $f$ can be defined on general grounds by
means of the subhalo mass,
$m(x) = 4\,\pi\,\rho_s\,r_s^3\,f(x)$, where $x=r/r_s$ -- for an NFW profile, it is simply
$f(x)=\ln (1+x)-x/(1+x)$. We have also defined $\Delta_x=\langle \rho \rangle_{r_x}/\rho_c$,
{\em i.e.} the ratio of the average subhalo density within a radius $r_x$ to the critical
density ($\Delta_{200}=200$). In the case of the pointlike Jacobi approximation corresponding
to the tidal definition of \citeeq{eq:rt_jacobi}, for instance, we have
\ben
\Delta_t(R) = 9\,M/(4\,\pi\,R^3\,\rho_c)\,,\nn
\een
where $M$ is the whole host galaxy mass.

The demonstration for the method of setting the tidal radius by equating the inner density to the
outer density, given in \citeeq{eq:rt_dens}, is trivial, and relies on the fact that the subhalo
scale density $\rho_s$, regardless of its profile and its mass, is only set by the concentration
parameter --- for an NFW profile, it reads
\ben
\rho_s = \frac{\Delta_{200}\,\rho_c}{3}\,\frac{c^3}{f(c)}\,.\nn
\een
If we write the density profile as $\rho(r)=\rho(x=r/r_s)=\rho_s \,u(x)$, then \citeeq{eq:rt_dens}
translates into $u(x) = \rho_{\rm tot}(R)/\rho_s$, which makes it clear again that $x_t$ depends
only on $c$ and $R$.

Finally, the cases of disk-shocking tidal effects are more subtle. In the differential
method, $x_t$ can readily be shown to be a function of $c$ and $R$ only from
\citeeq{eq:rt_diffds}. This is simply because the potential
$\tilde{\phi}(r_t)\propto r_s^2 g(x_t,c)$, where it is not necessary to specify function
$g$, while the kinetic energy
$\langle \epsilon_k\rangle (r_t)\propto r_t^2 \,\bar{g}(x_t,c,R)$, with the function $\bar{g}$ being
unspecified too, such that equating them leads to an equation that involves only the variables
$x_t$, $c$ and $R$. This proves that $x_t$ only depends only on $c$ and $R$.
The reasoning is similar for the so-called integrated disk-shocking methods, and also leads
to the dependence only on $R$ and $c$ of the associated $x_t$. Note that the independence
of $x_t$ on the subhalo mass cannot be read off \citefig{fig:diffds} nor off \citefig{fig:intds}
because subhalos with different masses in these plots have also different concentrations.
\subsubsection{Tidal disruption criterion and minimal concentration}
\label{sssec:disruption}

\begin{figure}[!t]
\centering
\includegraphics[width = 0.495\textwidth]{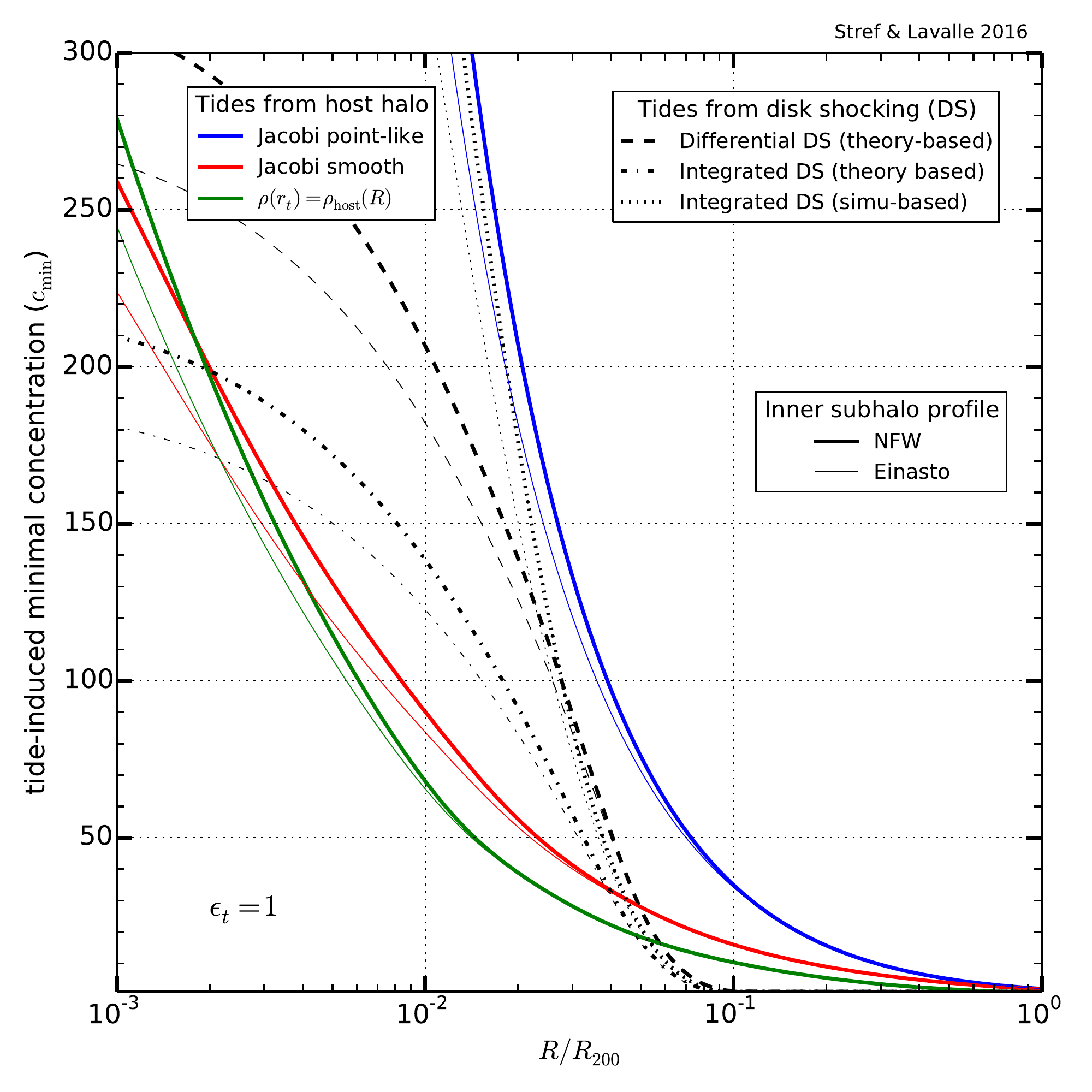}
\caption{\small Minimal concentration as a function of the
    dimensionless galactocentric radius $R/R_{200}$, induced by different tidal effects.
  The solid curves show the global tides effects discussed in \citesec{sssec:global_tides},
  and the non-solid curves show the disk-shocking effects discussed in
  \citesec{sssec:disk_tides}. See more comments in \citesec{sssec:disruption}.}
\label{fig:cmin}
\end{figure}

Equipped with several tidal radius definitions, we can now define a tidal disruption criterion by 
specifying the function $\zeta(r_t)$ introduced in \citeeq{eq:msub}, where $r_t$ is the
subhalo tidal radius. We remind the reader that the latter depends on all the specific subhalo
properties, and on its position in the host halo. In light of results obtained in
Ref.~\cite{Hayashi2003}, we may define the following very simple disruption function:
\ben
\zeta\left(x_t\equiv \frac{r_t}{r_s}\right) \equiv \theta\left(x_t - \varepsilon_t\right)\,,
\label{eq:disruption}
\een
where $\theta$ is the usual dimensionless step function, $r_s=r_s(m,c)$ is the subhalo scale 
radius, and the parameter $\varepsilon_t$ sets the minimal value allowed for $x_t$. This parameter
very likely depends on the inner subhalo density profile, and could also depend on the
specific process responsible for tidal stripping. Typical values found using dark-matter-only
simulations are $\varepsilon_t\approx 2$ (see Ref.~\cite{Hayashi2003}), but we may
wonder whether simulations can efficiently capture the continuous limit due to their limited
spatial/mass resolution. For definiteness, we will set $\varepsilon_t = 1$ in the following,
unless specified otherwise.

This translates into a minimal bound on the subhalo concentration, $c_{\rm min}(R)$,
as the surviving subhalos are only those with scale radii such that $r_t/r_s\geq \varepsilon_t$.
This concentration cutoff reads
\ben
c_{\rm min}(R) = \frac{\varepsilon_t}{\kappa} \,
\frac{r_{200}(m)}{r_t\left(c_{\rm min}(R),m,R\right)}\,,
\een
a transcendental equation that can be solved iteratively. Here, $\kappa=r_{-2}/r_s$ is fixed
by the choice of density profile ($\kappa=1$ for an NFW or an Einasto profile). In practice, we
will further impose that
\ben
c_{\rm min}(R) = {\rm Max}\left\{ c_{\rm min}(R) ; 1 \right\}\,.
\een
We emphasize that $c_{\rm min}$ does not actually depend on the subhalo mass, but only on its
location $R$ in the Galaxy. This is because $x_t$ is only a function of the concentration $c$ and
$R$, as explained in \citesec{sssec:xt}.

This concentration lower bound, $c_{\rm min}(R) $, is the very variable that differentiates
the tidal stripping methods discussed in \citesec{sssec:global_tides} and in
\citesec{sssec:disk_tides}. We report our calculations of $c_{\rm min}$ in \citefig{fig:cmin},
as a function of the dimensionless Galactic radius $R/R_{200}$ ($R_{200}=237$ kpc in the M11 model).
The curves related to the global tides are shown as solid colored lines, while those associated
with disk shocking are the nonsolid ones. Note that we have also included the baryons in the
calculation of the global tide effects (see \citesec{sssec:global_tides}). We also stress that we
preformed the calculation assuming two different inner subhalo profiles: an NFW profile (thick
curves), and an Einasto profile (thin curves) --- we took an index of $\alpha_e=0.17$ for the
latter. From the plot, there is no significant qualitative difference between these profiles,
except that Einasto subhalos are very slightly more resistant to gravitational tides.

We note that the most approximate method for the global tides, the pointlike Jacobi limit
given in \citeeq{eq:rt_jacobi}, is also the one that destroys subhalos most efficiently, even
more efficiently than disk shocking in the central parts of the Galaxy. It can therefore be used
for fast and conservative calculations, although it is highly sensitive to the estimate of the total
mass of the Galaxy, which is often ambiguous as it depends on the choice for the virial radius.
To make the discussion more quantitative, we recall that a $10^{-6}M_\odot$ subhalo has
a peak concentration of $\sim 60$, which will serve as a reference value here.
We see from the plot that the pointlike tide method affects such tiny objects
already from $20$ kpc and selects in only exponentially high concentration, while disrupting
less concentrated objects. This means that at the solar position all subhalos have already been
almost fully disrupted. The two other global tides methods [given in \citeeqss{eq:rt}{eq:rt_dens}],
which are much more realistic, give similar results and lead to much less tidal stripping than the
pointlike approximation. Subhalos of $10^{-6}M_\odot$ start to be strongly affected around 2-4 kpc
from the MW center in these scenarios.

Disk-shocking effects start to play a role only from $20$ kpc inward, as expected from the
typical gravitational size of the Galactic disk. All disk-shocking methods lead to more stripping
than global tidal effects, except for the pointlike approximation discussed above. Here again,
we see that the most approximate method, the integrated disk-shocking method fitted on
cosmological simulations and given in \citeeq{eq:intds_simu}, is the most efficient for destroying
or pruning subhalos. Besides being based on very crude approximations, we stress that it is also
likely biased by the resolution limit inherent to cosmological simulations, where only subhalos
with masses $\gtrsim 10^{4-7}$ can be
tracked. These massive objects are much less concentrated than their lighter brothers and sisters,
and more prone to stripping and disruption. In contrast, the less efficient method is the
one based on integrated disk shocking and given in \citeeq{eq:rt_intds}. Intermediate is the
method most motivated on theoretical grounds. Interestingly, the latter starts to deplete
subhalos of $10^{-6}M_\odot$ around the position of the Sun.

In summary, global tides tend to dominate the stripping beyond the disk, while disk shocking
dominates inward. This was obviously expected, but we quantified and illustrated these effects
rather exhaustively. Moreover, we showed that the pointlike Jacobi approximation makes it
irrelevant to include disk shocking, as it supersedes all other effects over the whole Galactic
range. Nevertheless, as we discussed above, this pointlike approximation is by far the worst
to make, while being conservative. Obviously, in a consistent and complete model, one has to
include all tides, those coming from global gravitational effects, and those coming from disk
shocking. This is what we will do when discussing our final results in \citesec{sec:res}.

\subsubsection{Tidal selection of the most concentrated objects: Shift of the average concentration}
\label{sssec:c_selection}
\begin{figure}[!t]
\centering
\includegraphics[width = 0.495\textwidth]{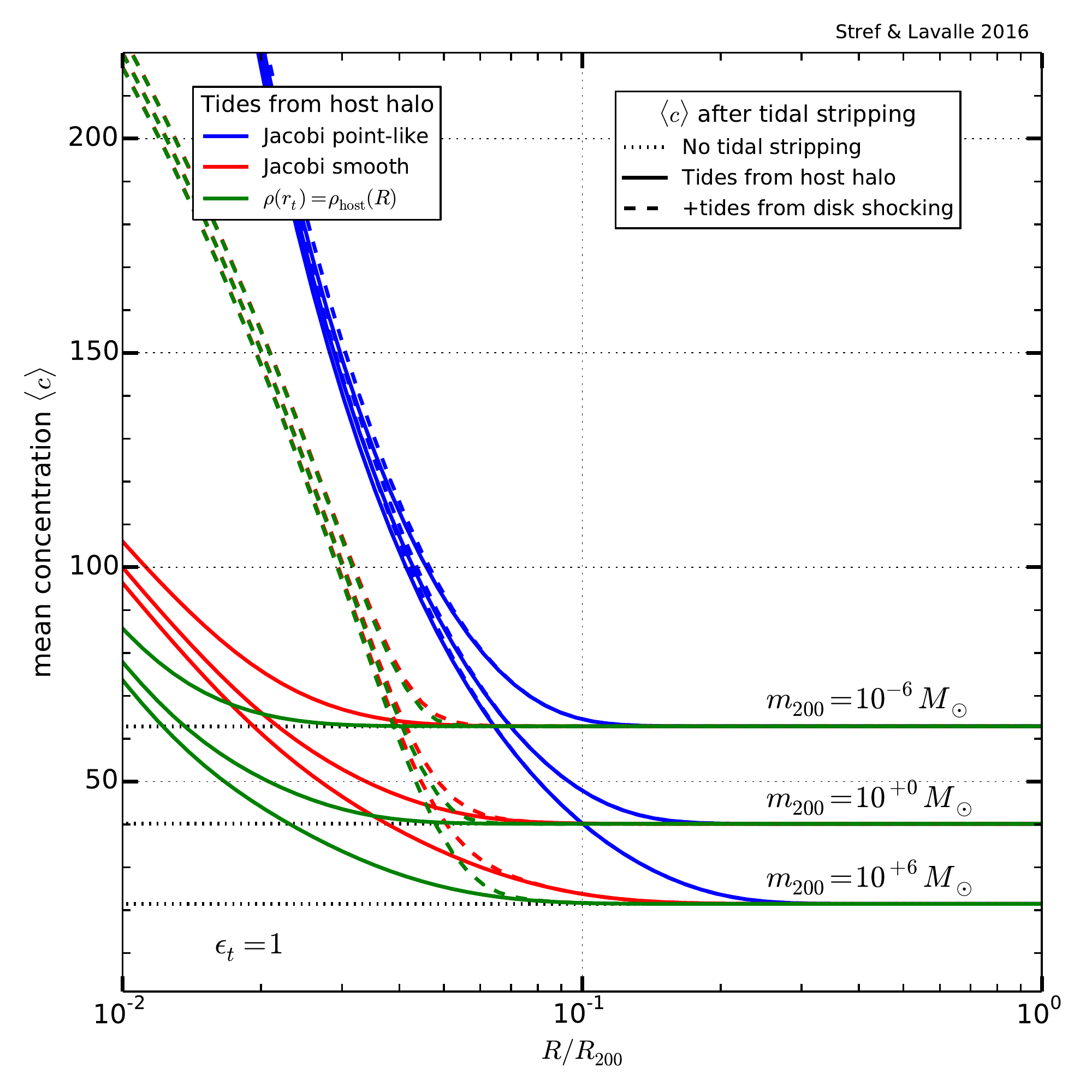}
\caption{\small Mean concentration as a function of the
    dimensionless Galactocentric radius $R/R_{200}$, induced by different tidal effects.
  The solid curves show the global-tides effects discussed in \citesec{sssec:global_tides},
  and the non-solid curves show the disk-shocking effects discussed in
  \citesec{sssec:disk_tides}. See more comments in \citesec{sssec:disruption}.}
\label{fig:cmean}
\end{figure}
By depleting the lower tail of the concentration distribution, tidal effects modify the average
concentration of subhalos as a function of their mass. This can be explicitly calculated
by means of the first moment of the concentration function, given in \citeeq{eq:cmean}.
The increase in the average concentration merely comes from the fact that tidal effects reduce the
concentration range from below by $c_{\rm in}(R)\geq 1$. In reality, the concentration
function should not be truncated that sharply, but this truncation still captures the main
physical effects at play.

We illustrate this in \citefig{fig:cmean}, where we report our calculations of
$\langle c(m)\rangle$ as a function of the dimensionless Galactic radius $X_{200}=R/R_{200}$ for
three different subhalo masses, $10^{-6}$, $1$, and $10^6M_\odot$, and for all the tidal-stripping
methods introduced above. The asymptotic values of $\langle c(m)\rangle$ at $X_{200}\to 1$
correspond to the average concentration computed in the range $c_{\rm min}=1$, $c_{\rm max}=\infty$.
As we go inward, tidal effects come into play and $c_{\rm min}$ increases, leading to the
increase in $\langle c(m)\rangle$. Recalling that the concentration function is Gaussianly
suppressed beyond the median value $c_0(m)\approx$10-100, in the considered subhalo mass range,
we can therefore read off from the plot that most of subhalos with masses larger than that of a
given curve are tidally depleted as the curve exceeds $\sim 100$. This trend is consistent with
previous studies performed from dark-matter-only cosmological simulation results (see
\eg~Refs.~\cite{Hayashi2003,vandenBoschEtAl2005,Diemand2007a,Kamionkowski2010,Moline2016,Jiang2016}), or
from simple analytic approximations (see \eg~Ref.~\cite{Bartels2015}), but these works did not
include baryonic effects. Here we provide quantitative estimates for both baryonic and dark matter
tidal effects, and comparisons between different approaches.

\subsubsection{Impact of tidal effects on the calibration and normalization procedure}
\label{sssec:norm_tides}
It may prove useful to summarize the way tidal effects are integrated in the full procedure
in practice. As discussed in \citesec{sssec:norm_sub}, we calibrate the subhalo population
by considering only the so-called global tidal effects presented in \citesec{sssec:global_tides}.
These global tidal effects translate into a function $c_{\rm min}^0(R)$ that cuts
the concentration PDF from below and allows us to determine both $N_{\rm sub}^0$ and
the associated normalization of the whole subhalo phase space $K_0$. This must be done
without baryons at all, consistently with the fact that the calibration is based upon
dark-matter-only simulation results. Then, we compute the final phase-space normalization $K$
that accounts for the baryonic tides (both the global tide and the disk-shocking calculations),
which are characterized by a new cutoff function $c_{\rm min}(R)$. We obtain the final number
$N_{\rm sub}$ of subhalos by demanding that the overall subhalo mass
density be unaffected at very large radii, far from the disk, where baryonic effects can be
neglected. This can be rephrased as setting $N_{\rm sub} = (K/K_0)\,N_{\rm sub}^0$.

\subsection{Reference Galactic halo model (including a subhalo population)}
\label{ssec:ref_mod}
Before discussing in detail the observables relevant to DM searches in the next
  section, we define here our reference Galactic model:
  \bi
\item[1] Our reference Galactic mass model, which fixes both the global dark halo (including
  subhalos) and the Galactic baryonic content is the M11 model (see \citesec{ssec:dynamics}).
\item[2] Global tides induced by the global mass distribution are by default calculated
  in the smooth Jacobi limit [see \citesec{sssec:global_tides} and \citeeq{eq:rt}].
\item[3] Tides induced by disk shocking will be calculated by default with the differential
  disk-shocking method [see \citesec{sssec:disk_tides} and \citeeq{eq:rt_diffds_it}].
\item[4] The default tidal disruption efficiency will be set to $\varepsilon_t=1$
  [see \citesec{sssec:disruption} and \citeeq{eq:disruption}].
  \ei
  Any result in the following will derive from this default configuration unless specified
  otherwise.
  We will also consider two benchmark cases for the initial mass index, $\alpha_M=1.9$ and
  $\alpha_M=2$. Similarly, we will consider two benchmark cutoff subhalo masses,
  $m_{\rm min}=10^{-6}\,M_\odot$ and $m_{\rm min}=10^{-10}\,M_\odot$.
\section{Concrete results: Mass profiles, number density profiles, luminosity profiles,
  and boost factors}
\label{sec:res}
In the previous section, we introduced the whole scheme to derive a Galactic subhalo
population consistently with current dynamical constraints, assuming only a smooth-halo component
and spherical symmetry. This scheme was integrated in a C++ numerical code, and in this section,
we will present our main results. In
\citesec{ssec:mprofiles}, we will first inspect the overall subhalo mass profile that comes out
as a result of our tidal-stripping procedure. Then, in \citesec{ssec:nsub}, we will present
  the corresponding subhalo number density profile. Finally, in \citesec{ssec:boosts}, we will
show how the obtained subhalo distribution translates into an annihilation profile, and
subsequently quantify the associated annihilation boost factor. Throughout this section,
we will also comment on the specific impacts of the initial
subhalo mass distribution index $\alpha_M$ and the minimal subhalo mass (see \citesec{sssec:dPdm}).
\subsection{Mass profiles}
\label{ssec:mprofiles}
\begin{figure*}[!t]
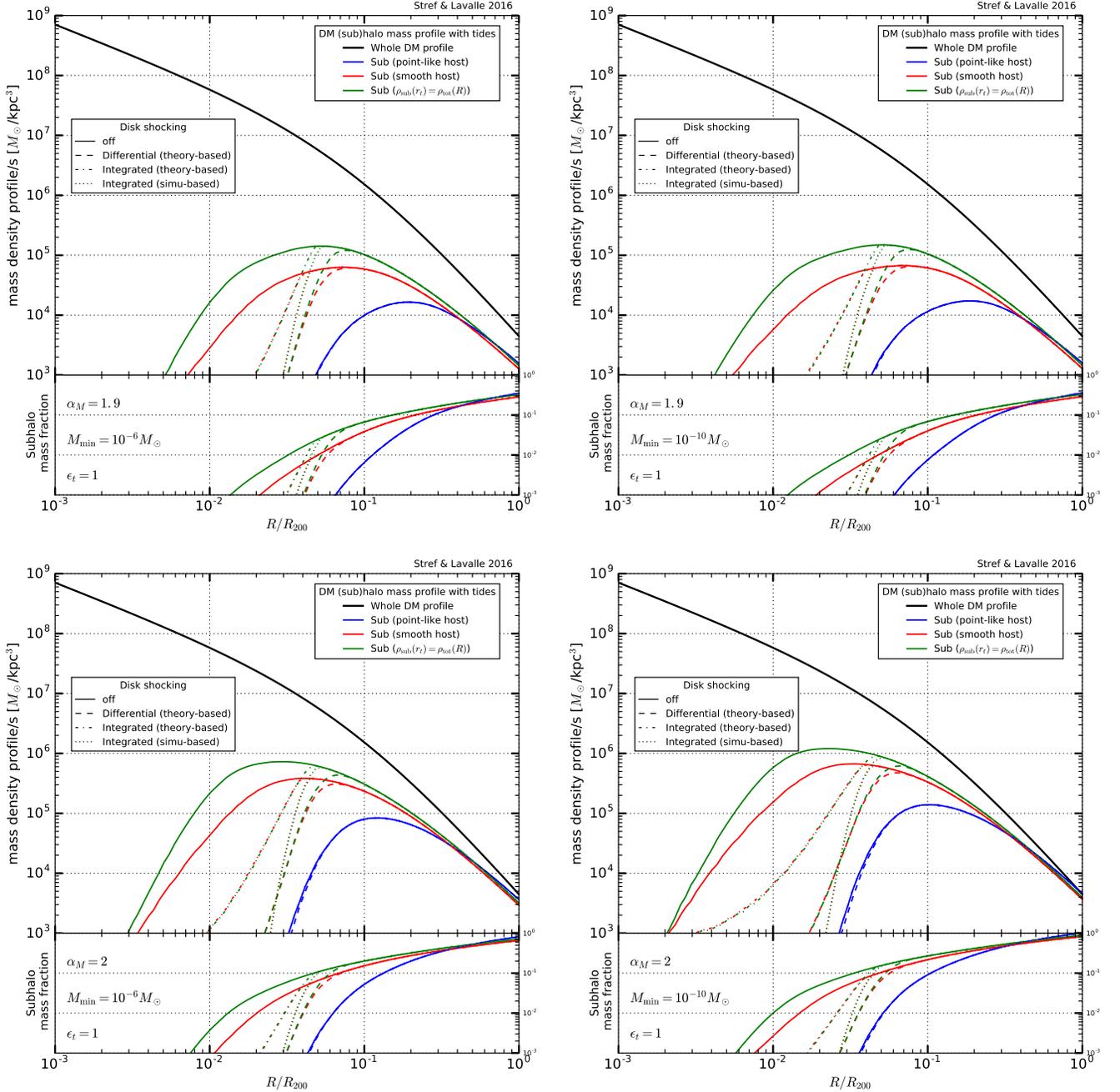

\centering
\includegraphics[width = 0.495\textwidth]{{{fig_density_profiles_tides_residuals_alpha1.9_mmin1e-06_epstide1}}}
\includegraphics[width = 0.495\textwidth]{{{fig_density_profiles_tides_residuals_alpha1.9_mmin1e-10_epstide1}}}
\includegraphics[width = 0.495\textwidth]{{{fig_density_profiles_tides_residuals_alpha2_mmin1e-06_epstide1}}}
\includegraphics[width = 0.495\textwidth]{{{fig_density_profiles_tides_residuals_alpha2_mmin1e-10_epstide1}}}
\caption{\small Mass profiles obtained for different subhalo configurations as embedded
    in the M11 Galactic mass model. {\bf Top (bottom)} panels illustrate our results for a mass
    index of $\alpha_M= 1.9$ (2, respectively). {\bf Left (right)} panels adopt a minimal subhalo
    mass of $m_{\rm min}=10^{-6}\,M_\odot$ ($10^{-10}\,M_\odot$, respectively).
    {\bf All panels}: the black solid
    curve shows the constrained overall dark halo mass density profile, and the other curves
    feature the global contribution of subhalos for different tidal effect calculations. Blue, red,
    and green curves illustrate the global tides induced by the dark halo itself calculated
    respectively in the pointlike Jacobi limit, the smooth Jacobi limit, and the density equality
    limit discussed in the text (see \citesec{sssec:global_tides}). Dashed, dashed-dotted and
    dotted curves show additional impact of the gravitational stripping induced by disk shocking
    respectively based on the differential method, the integrated method, and the empirical fit
    done on simulation results; these methods are discussed in \citesec{sssec:disk_tides}.}
\label{fig:mprofiles}
\end{figure*}
\begin{table*}[t]
  \begin{tabular}{| c || c | c | c | c |}
    \hline
    Mass function index & Total number & Phase-space normalization & Fraction in  & Total mass fraction \\
    $\alpha_M$ & of surviving subhalos & $K_w$ & local density (average) & within $R_{200}$ \\
    \hline
    $\alpha_M=1.9$ & $5.19\times 10^{18}$  & 0.9638 &  0.04\% & 14.69\% \\
    \hline
    $\alpha_M=2 $  & $2.84\times 10^{20}$ & 0.9639 & 0.84\% & 47.88\% \\
    \hline
  \end{tabular}
  \caption{\small Results for our subhalo model as embedded in the M11 Galactic configuration, when
    all tidal effects are included (see reference model configuration in
      \citesec{ssec:ref_mod}). Here, we take a cutoff subhalo mass of $10^{-10}M_\odot$, and a
    tidal disruption efficiency of $\varepsilon_t=1$.}
  \label{tab:check}
\end{table*}
In \citesec{ssec:halo_and_sub}, we have introduced the smooth and subhalo dark matter components
from the overall density profile $\rho_{\rm tot}$ in \citeeq{eq:rhotot}, which we rewrite with
an explicit spherical symmetry to make the present discussion clear:
\ben
\rho_{\rm tot}(R) = \rho_{\rm sm}(R) + \rho_{\rm sub}(R)\,.
\een
\begin{widetext}
  We remind the reader that $\rho_{\rm tot}$ is subject to dynamical constraints, and we have adopted
  the M11
model as a template Galactic mass model (see \citesec{ssec:dynamics}). The smooth dark
matter component $\rho_{\rm sm}$ featured above can only be derived {\em a posteriori}, after
having determined the subhalo component $\rho_{\rm sub}$ given in
\citeeq{eq:def_rhosub}. Making the tidal cutoff $c_{\rm min}(R)$ explicit, the latter
reads
\ben
\rho_{\rm sub}(R) &=&
\frac{N_{\rm sub}}{K_w} \,\frac{d{\cal P}_V(R)}{dV}\,
\int_{m_{\rm min}}^{m_{\rm max}} dm 
\int_{c_{\rm min}(R)}^{c_{\rm max}} dc \,
m_t(r_t(c,m,R),m,c)\,\frac{d{\cal P}_m}{dm}\,\frac{d{\cal P}_c}{dc}\nn\\
\text{with}\;K_w &=& 4\,\pi\int_0^{R_{200}}dR\,R^2\,\frac{d{\cal P}_V(R)}{dV}
\int_{m_{\rm min}}^{m_{\rm max}} dm 
\int_{c_{\rm min}(R)}^{c_{\rm max}} dc \,
\frac{d{\cal P}_m}{dm}\,\frac{d{\cal P}_c}{dc}\,,
\een
where $N_{\rm sub}$ is the total number of subhalos, $m_t$ is the subhalo mass contained in the
tidal radius $r_t$, and the ${\cal P}$'s define the global subhalo phase space, normalized to
unity thanks to $K_w$, introduced in \citesec{sssec:overall_sub}.
\end{widetext}
Taking the M11 Galactic mass model as a reference, the associated prediction of the overall
subhalo mass density profile is shown in Fig.~\ref{fig:mprofiles}, where we also represent the
impact of the mass index ($\alpha_M=1.9/2$ in the top/bottom row's panels) and that of the minimal
subhalo mass ($10^{-6}/10^{-10}M_\odot$ in the left/right column's panels). In each panel, we give
predictions for all the methods introduced in \citesec{ssec:tides} to compute the tidal stripping
and associated subhalo disruption -- the upper part of each plot shows the mass density profile,
and the lower part shows the subhalo mass fraction, as functions of the dimensionless Galactic
radius $R/R_{200}$. Subhalo mass profiles relying only on global tides are reported as solid colored
curves -- see \citesec{sssec:global_tides} -- while
those incorporating disk-shocking effects as well are shown as dashed (differential), dash-dotted
(integrated), and dotted (simulation-fit inspired method) -- see
\citesec{sssec:disk_tides}. We recall that our reference model is based on the global tides
evaluated in the smooth Jacobi limit (dubbed {\em smooth host} in the plots), given in
\citeeq{eq:rt}, and on the differential disk-shocking method given in \citeeq{eq:rt_diffds_it}.
It is represented as dashed red curves in each panel. Some illustrative numbers
can also be found in \citetab{tab:check}.

Overall, these results show that tidal effects strongly deplete the subhalo population in the
central parts of the Galaxy, and underline the impact of disk shocking, which plays an important
role. This gives rise to a cored-like spatial distribution inward, before the full disruption of
subhalos in the very center ($\lesssim 4$ kpc). These are generic features observed in cosmological
simulations, but our analytic procedure allows us to make predictions down to much lower spatial
and mass scales, in a dynamically constrained and consistent frame. Going to more specific
global tidal stripping configurations, we see that the global pointlike Jacobi method,
which is clearly too approximate as it does not account for the host halo and subhalo profiles'
details, disrupts almost all subhalos within
$R/R_{200}\lesssim 0.1 \Leftrightarrow R\lesssim 20$ kpc, making disk-shocking effects even
irrelevant. This strongly suppresses the local ($R/R_{200}\sim 0.3$-0.4) subhalo mass fraction,
typically to $\ll 1\%$. The two other more physically motivated global methods provide
slightly more optimistic predictions, with a local subhalo mass fraction $\sim 10\%$. When
disk-shocking effects
are further included, however, it decreases down to $\lesssim 1\%$. Still, we will see
in \citesec{ssec:boosts} that this low fraction is somewhat compensated for, in terms of
annihilation rate, by a tidal selection of more concentrated objects. The impact of the
minimal subhalo mass is only noticeable for a mass index of 2, as most of the mass fraction
is then carried by the smallest objects, which are much more resilient to tidal effects.
Since the minimal subhalo mass can in principle be determined by the interaction properties
of WIMPs (more or less straightforwardly related to its mass), most of the theoretical
uncertainties are then featured by $\alpha_M$. For our reference model, we see that while
the mass fraction can vary by a factor of $\sim 2$ between $\alpha_M=1.9$ and $\alpha_M=2$
in the outskirts of the Galaxy, its differential value is much more sensitive because of more
efficient tidal selection of lighter subhalos. In the latter case, the variation can reach an
order of magnitude. This should have an impact on predictions for direct subhalo searches
(see \eg\ Refs.~\cite{Feldmann2015,ClarkEtAl2016}).

Finally, we note that the amount of subhalo mass lost during disk crossings could in principle be
quantified from our method, which may be used to size the impact of the smallest pruned subhalos on
the high tail of the WIMP velocity distribution (see \eg\ Refs.~\cite{Zemp2009,Kuhlen2010}). Indeed,
disk shocking induces a net kinetic energy gain for the pruned WIMPs. This,
however, goes beyond the scope of this work.

\subsection{Number density profiles}
\label{ssec:nsub}
Another interesting observable complementary to the mass profile discussed above is the
number density profile, which will provide another angle to understand the role of tides and
their immediate consequences. In \citeeq{eq:def_nsub}, we provided the differential subhalo
number density as a function of the cosmological mass $m=m_{200}$, which is not the actual
tidal subhalo mass $m_t$. It is easy to show that $m$ and $m_t$ are related by 
\ben
m_t = m\,\Delta(\vec{x},c)\,,
\een
where function $\Delta(\vec{x},c)\leq 1$ encodes the dependence of $m_t$ (or $r_t$) on position
and concentration. Since this dependence factorizes, one can merely trade the phase-space volume
$dV\,dm\,dc$ for $dV\,dm_t\,dc$ by means of the classical Jacobian. This is equivalent
to the projection of $dn_{\rm sub}/dm$ on $dn_{\rm sub}/dm_t$ directly by means of the delta function
$\delta(m_t-m_{200}\,\Delta(\vec{x},c))$. Note that physically, different masses $m$ and different
concentrations $c$ will lead to to the same tidal mass $m_t$, which means that the cosmological
nature of the initial mass spectrum is somewhat erased by tidal stripping.

\begin{widetext}
  We get
  \ben
\label{eq:dndmt}
  \frac{dn_{\rm sub}(\vec{x},m_t)}{dm_t} =
      \frac{N_{\rm sub}}{K_w} \, \frac{d{\cal P}_V(\vec{x})}{dV}\,
         \int_{m_{\rm min}}^{m_{\rm max}}dm\, 
         \int_{c_{\rm min}(\vec{x})}^{c_{\rm max}} dc \,\frac{d{\cal P}_c}{dc}\,
         \big|{\cal J}_{m_t/m}\big|\,\frac{d{\cal P}_m(m)}{dm}
         \,\delta(m-m_t/\Delta(\vec{x},c)) \,,
  \een
  where the Jacobian $\big|{\cal J}_{m_t/m}\big|=\Delta(\vec{x},c)$, and where the dependence on
  position $\vec{x}$ amounts in our study to a dependence in radial position $R$.

It is straightforward to verify that
\ben
n_{\rm sub}(\vec{x}) = \int_{m_{\rm min}}^{m_{\rm max}} dm \frac{dn_{\rm sub}(\vec{x},m)}{dm}
= \int_{m_{t,{\rm min}}}^{m_{t,{\rm max}}}dm_t \frac{dn_{\rm sub}(\vec{x},m_t)}{dm_t}\,,
\label{eq:dndmt_dndm}
\een
which shows the consistency of \citeeq{eq:dndmt} with \citeeq{eq:def_nsub}. In the equation
above, the maximal tidal mass $m_{t,{\rm max}}$ can be set to $m_{\rm max}$ while the minimal tidal
mass $m_{t,{\rm min}}$ can readily be obtained from $m_{\rm min}$:
\ben
m_{t,{\rm min}}(\vec{x}) = m_{\rm min}\,\Delta(\vec{x},c_{\rm min}(\vec{x}))\leq m_{\rm min} \,.
\een
The differential number density profile $dn_{\rm sub}/dm_t$ is actually proportional to a local
mass distribution that we can defined as a locally normalized function
\ben
\frac{d\widetilde{\cal P}_{m}(m,\vec{x})}{dm} &=&
\frac{d{\cal P}_m(m)}{dm} \frac{\int_{c_{\rm min}(\vec{x})}^{c_{\rm max}} dc \,\frac{d{\cal P}_c}{dc}}
     {\int_{m_{\rm min}}^{m_{\rm max}}dm\, \frac{d{\cal P}_m(m)}{dm}\,
         \int_{c_{\rm min}(\vec{x})}^{c_{\rm max}} dc \,\frac{d{\cal P}_c}{dc}}\nn\\
\frac{d\widetilde{\cal P}_{m_t}(m_t,\vec{x})}{dm_t} &=&
\frac{\int_{c_{\rm min}(\vec{x})}^{c_{\rm max}} dc \,\frac{d{\cal P}_c}{dc}\,
         \big|{\cal J}_{m_t/m}\big|\,\frac{d{\cal P}_m(m)}{dm}
         \,\delta(m-m_t/\Delta(\vec{x},c))}
     {\int_{m_{\rm min}}^{m_{\rm max}}dm\, 
         \int_{c_{\rm min}(\vec{x})}^{c_{\rm max}} dc \,\frac{d{\cal P}_c}{dc}\,
         \big|{\cal J}_{m_t/m}\big|\,\frac{d{\cal P}_m(m)}{dm}
         \,\delta(m-m_t/\Delta(\vec{x},c)) }\,,
     \label{eq:dpeffdm}
\een
where we have given both the virial $d\widetilde{\cal P}_{m}/dm$ and tidal
$d\widetilde{\cal P}_{m_t}/dm_t$ functions, and where the numerators ensure the local normalization
to unity. We see from the above equation that the spatial-dependence of the range in concentration
induced by $c_{\rm min}(\vec{x})$ will modify the mass index $\alpha_M$ which characterizes
the initial mass function $d{\cal P}_m/dm$ given in \citeeq{eq:dpdm}, leading to an effective
mass index $\widetilde{\alpha}_M(\vec{x})\geq \alpha_M$, because the more massive subhalos will be
more efficiently disrupted toward the central parts of the Galaxy.
\end{widetext}
This is illustrated in \citefig{fig:dndm}, where we report $m^2 dn_{\rm sub}/dm$ (in terms of both
the cosmological and tidal masses) at different positions: inside the disk ($R=8$ kpc), at the
very end of the disk ($R=20$ kpc), and far from the disk ($R=100$ kpc). The initial mass index
has been fixed to $\alpha_M=2$, but the qualitative behavior would be the same with 1.9.
The initial mass distribution is explicitly reflected in the dotted curves, which show the
differential number density before tidal stripping is applied, and which are flat in
the graph because of the $m^2$ rescaling (here, we took the same total number of subhalos as for
the global tides-only case). There is a sharp cutoff corresponding to the minimal
subhalo mass ($10^{-10}/10^{-6}\,M_\odot$ in the left/right panel). The difference in amplitude
between these dotted curves only comes from the difference in the spatial distribution value,
which initially scales like $\rho_{\rm tot}$ (the amplitude increases as the radius $R$
decreases). As tidal effects are plugged in, the dotted curves are converted to the dashed
curves, still in terms of the cosmological mass $m=m_{200}$. While these curves are close to
each other at $R=100$ kpc (black curves), we see a departure of the tidal curve below
intermediate masses at $R=20$ kpc (red curves), and finally a very strong departure over the
whole mass range at $R=8$ kpc (blue curves). The latter exhibits a local effective mass
index $\widetilde{\alpha}_M\sim 3 \gg \alpha_M$, and a strongly depleted amplitude, such
that despite the initially much larger number density at $R=8$ kpc, the hierarchy is now
reverted. This reflects the impact of tidal disruption whose efficiency is strongly enhanced
in the disk because of disk shocking. We stress that this only illustrates the behavior of the
{\em local} effective mass index. When integrated over the whole halo, the effect would still
be visible but much less pronounced because of the significant weight carried by subhalos
located far from the disk, leading to a {\em global} effective mass index $\gtrsim \alpha_M$, as
obtained in other studies (see \eg~\cite{Jiang2016,vandenBosch2016}). Still, it is striking
to see how tidal effects deeply affect the phase space locally.

So far, we discussed \citefig{fig:dndm} in terms of cosmological mass. If we now concentrate
on the differential subhalo number density in terms of real (tidal) mass $m_t$ (solid curves),
the picture gets modified. The most noticeable difference comes from the fact that the boundary
is less sharp at low masses, and that this boundary does not coincide with $m_{\rm min}$ anymore,
but can be much lower (depending on the tidal disruption parameter $\varepsilon_t$ -- see
\citesec{sssec:disruption}).
If fact, only subhalos with very large concentrations will have $m_t\sim m$, so subhalos located
at a given position on the cosmological-mass curve will migrate toward the left part of the
tidal-mass curve as their concentration decreases. Therefore, $m_t$ no longer carries
cosmological information by itself, only the internal structure of subhalos does. Moreover
we emphasize that only $dn_{\rm sub}/dm_t$ is a directly observable quantity, not
$dn_{\rm sub}/dm$. Though not obvious from the graphs, the differential
number density curves reported in \citefig{fig:dndm} do obey the consistency relation given in
\citeeq{eq:dndmt_dndm}.

Finally, if we perform the mass integral of $dn_{\rm sub}/dm$, we can get the subhalo number
density profile $n_{\rm sub}(r)$ in the Milky Way [see \citeeq{eq:dndmt_dndm}]. We illustrate
this in \citefig{fig:nprofile}, where we report results for different configurations, all
assuming the M11 Galactic mass model: without tidal effects (dotted curves), with global
tides only (dashed curves), with global tides and disk shocking (solid curves); for two
different values of the mass index $\alpha_M$ (1.9/2 for the green/blue curves), and two
values for the minimal cutoff mass ($m_{\rm min}=10^{-6}\,M_\odot/10^{-10}\,M_\odot$ for the
left/right panel).
For tidal effects, we used the reference modeling summarized in \citesec{ssec:ref_mod}.
The cases with tidal effects do not strictly converge toward the case without tides because
the latter comes with a slightly smaller total number of subhalos as a normalization bias
(it is calculated by requiring the same mass fraction in a reference range as in the global
tides-only case, though without stripping or disruption allowed, hence more subhalos --
see \citesec{sssec:norm_sub}). The difference induced by $m_{\rm min}$ is simply in the relative
amplitude of the number density, as explained in more detail in \citesec{ssec:mmin}.

\begin{figure*}[!t]
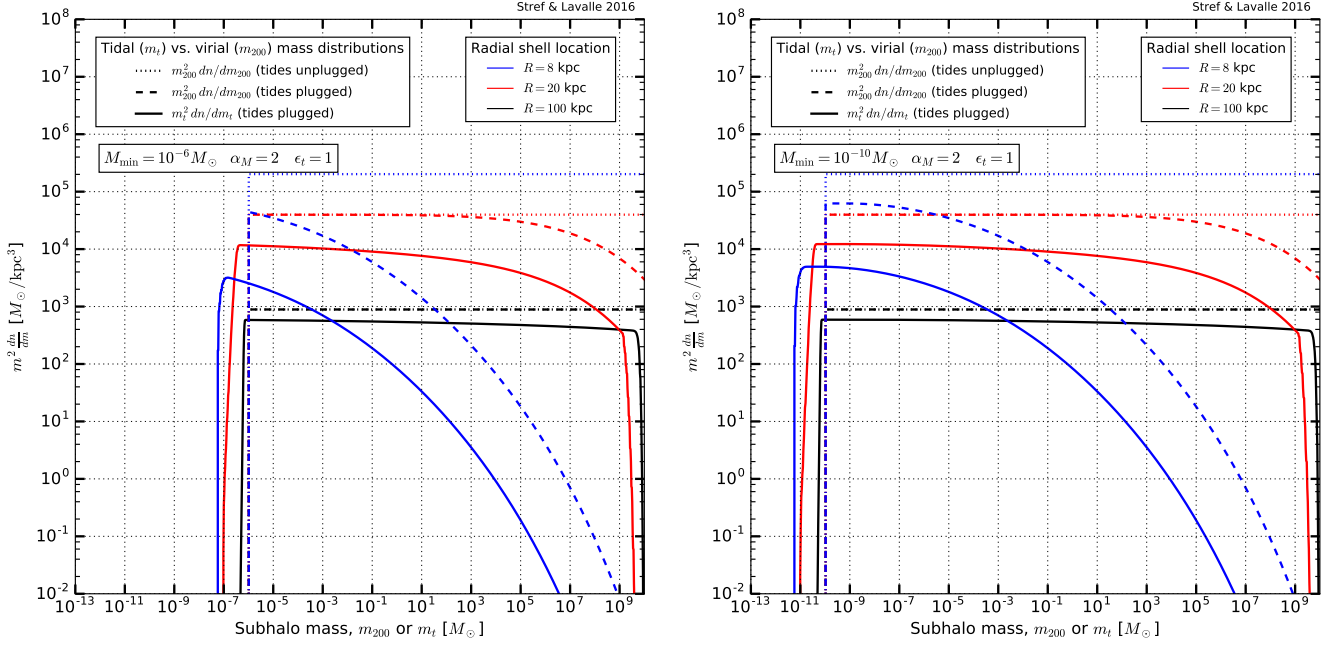

\centering
\includegraphics[width = 0.495\textwidth]{{{fig_dndm_mmin1e-06}}}
\includegraphics[width = 0.495\textwidth]{{{fig_dndm_mmin1e-10}}}
\caption{\small \textbf{Left panel:} mass-differential number density in terms of both
    the cosmological mass $dn_{\rm sub}/dm$ (dotted/dashed curves for tidal effects
    unplugged/plugged) and the real tidal mass $dn_{\rm sub}/dm$ (solid curves), assuming a cutoff
    mass $m_{\rm min}=10^{-6}\,M_\odot$, and calculated at 3 positions
    ($R=8/20/100$ kpc in blue/red/black), as a function of the relevant
    mass (a scaling of $m^2$ is applied as $\alpha_M=2$ is chosen here). \textbf{Right panel:}
    same as in the left panel but with $m_{\rm min}=10^{-10}\,M_\odot$.}
\label{fig:dndm}
\end{figure*}

\begin{figure*}[!t]
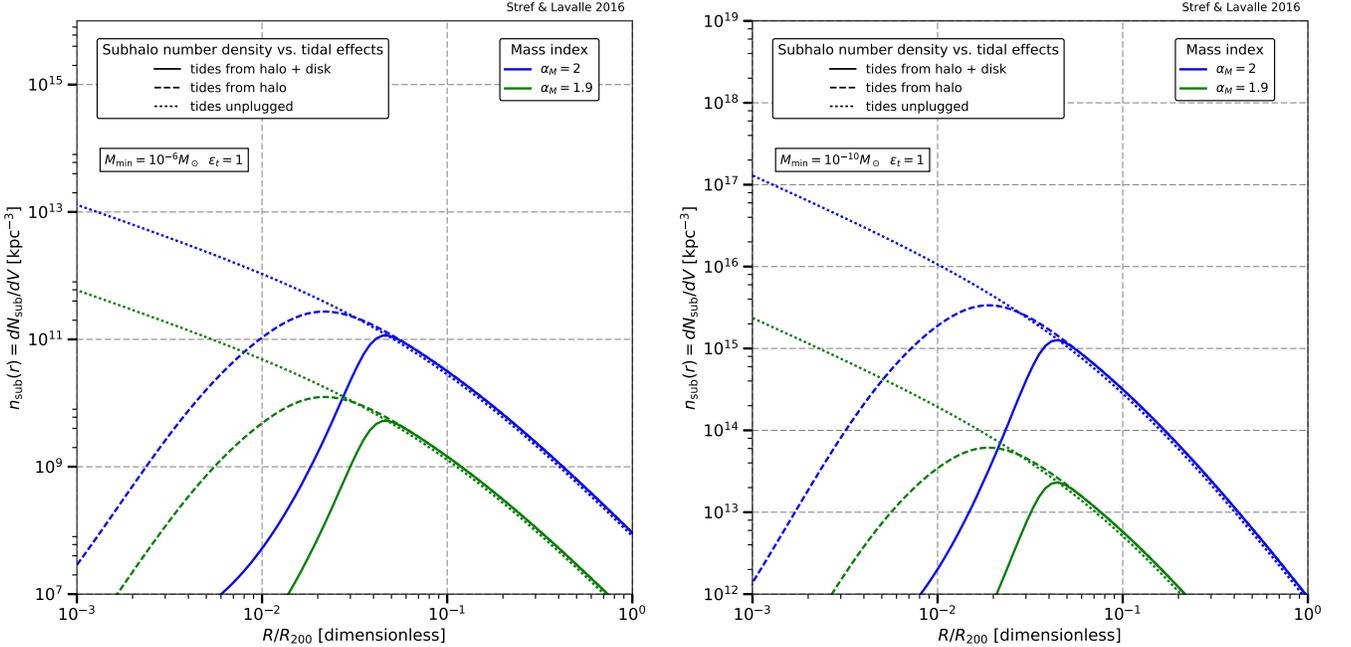

\centering
\includegraphics[width = 0.495\textwidth]{{{fig_number_density_profiles_tides_mmin1e-06_epstide1}}}
\includegraphics[width = 0.495\textwidth]{{{fig_number_density_profiles_tides_mmin1e-10_epstide1}}}
\caption{\small \textbf{Left panel:} Subhalo number density profile assuming the M11
    Galactic mass model (dotted/dashed/solid curves for tidal effects unplugged/halo/halo+disk from
    the reference model -- see \citesec{ssec:ref_mod}), two different cases for the mass index
    $\alpha_M$ (1.9/2 for the green/blue curves), and assuming a cutoff
    mass $m_{\rm min}=10^{-6}\,M_\odot$. \textbf{Right panel:}
    same as in the left panel but with $m_{\rm min}=10^{-10}\,M_\odot$.}
\label{fig:nprofile}
\end{figure*}

\subsection{Annihilation rate profiles and boost factors}
\label{ssec:boosts}
\begin{figure*}[!t]
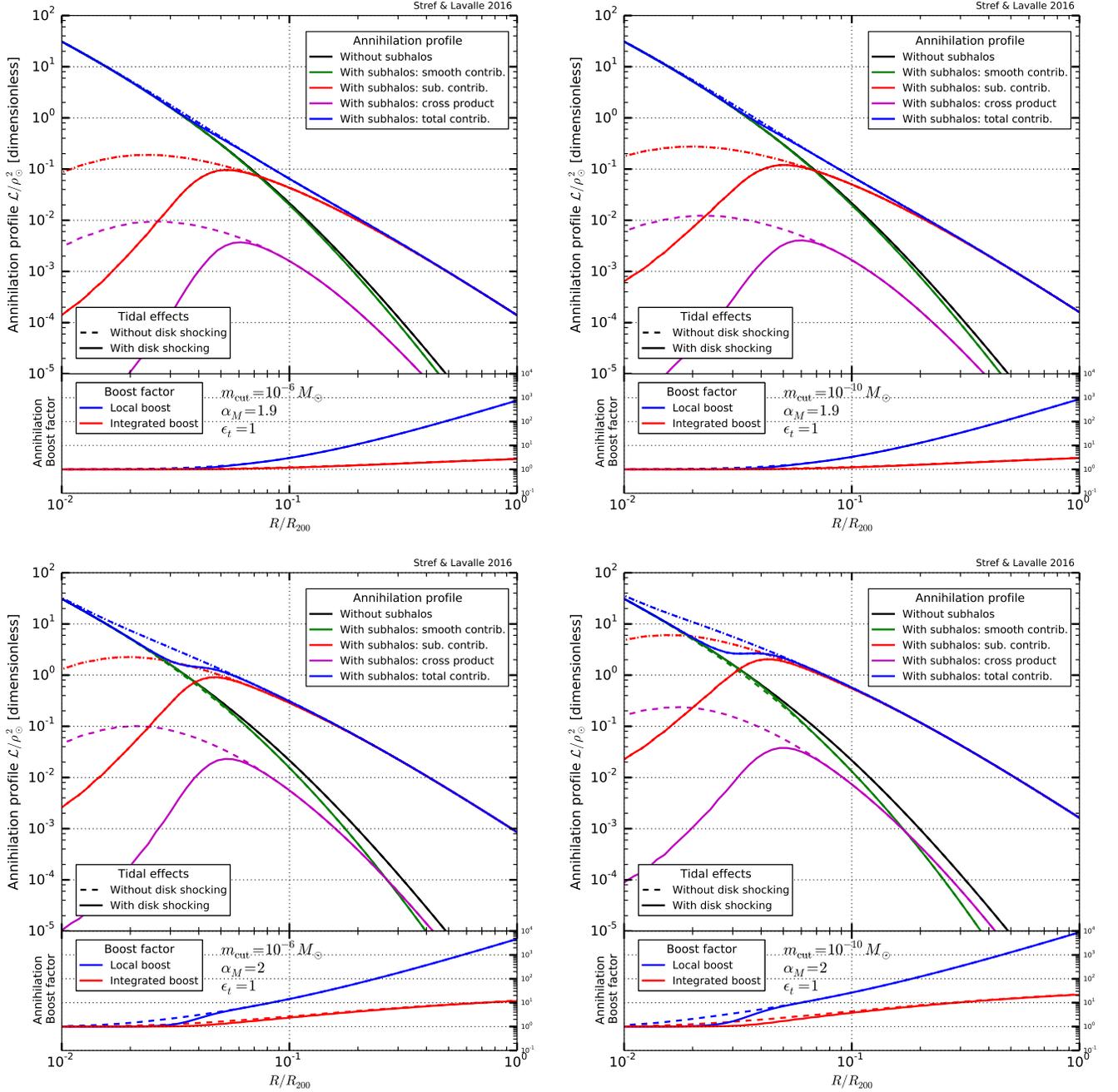

\centering
\includegraphics[width = 0.495\textwidth]{{{fig_annihilation_profile_smoothmod16_mmin1e-06_alpha1.9_eps1}}}
\includegraphics[width = 0.495\textwidth]{{{fig_annihilation_profile_smoothmod16_mmin1e-10_alpha1.9_eps1}}}
\includegraphics[width = 0.495\textwidth]{{{fig_annihilation_profile_smoothmod16_mmin1e-06_alpha2_eps1}}}
\includegraphics[width = 0.495\textwidth]{{{fig_annihilation_profile_smoothmod16_mmin1e-10_alpha2_eps1}}}
\caption{\small Annihilation/luminosity profiles for different assumptions. {\bf Upper part of each
  panel}:
  Black curves show the profile when neglecting subhalos, blue curves show the overall
  profile when subhalos are included and all tidal effects are considered, green curves show the
  separate contribution of the smooth halo, red curves show the contribution of subhalos, and
  magenta curves the contribution of the subhalo-smooth halo cross product. Dashed lines show the
  impact
  of neglecting disk-shocking effects. {\bf Lower part of each panel}: Differential (blue curves)
  and integrated (red curves) boost factors. {\bf Upper/lower row}: $\alpha_M=1.9/2$.
  {\bf Left/right panels:} Minimal subhalo mass of $10^{-6}/10^{-10}M_\odot$.}
\label{fig:annprofiles}
\end{figure*}
In this section, we discuss the potential enhancement the presence of dark matter subhalos
may induce in the WIMP annihilation rate in the Galaxy, usually dubbed the {\em boost factor}.
Here, we will only determine the differential and integrated boost factors on the annihilation
rate, not on the observable cosmic-ray or gamma-ray fluxes. We remind the reader that in terms of
these fluxes, the annihilation boost factor is angular dependent for gamma rays
\cite{Bergstroem1999a}, while it is energy dependent for antimatter cosmic rays
\cite{Lavalle2007a,Lavalle2008c}.

In this work, we will assume that subhalos do not superimpose, such that we will not account
for the potential existence of sub-subhalos, which might be relevant in the most massive
subhalos, as they have formed at later epochs than the lightest ones. We still stress that
any inclusion of sub-subhalos should be consistent with the normalization and calibration
procedures one subscribes to (in particular, the overall mass function should be recovered
after all layers of subhalos have been accounted for).
\begin{widetext}
With this assumption, the total
annihilation rate can be derived starting from a discrete distribution of $N_{\rm sub}$ subhalos,
where the total dark matter density would be given by
\ben
\rho_{{\rm tot},N_{\rm sub}}(\vec{x}) = \rho_{\rm sm}(\vec{x})+
\sum_i^{N_{\rm sub}}M_{t,i}\delta(\vec{x}-\vec{x}_i)\,,
\een
where each subhalo $i$ is pointlike and allocated a tidal mass $M_{t,i}$, and $\rho_{\rm sm}$ is
the smooth dark matter density. Squaring this equation, we get
\ben
\rho_{{\rm tot},N_{\rm sub}}^2(\vec{x}) = \rho_{\rm sm}^2(\vec{x})+
\sum_{i\geq j}^{N_{\rm sub}}M_{t,i}\delta(\vec{x}-\vec{x}_i)\,M_{t,j}\delta(\vec{x}-\vec{x}_j)+
2\,\rho_{\rm sm}(\vec{x})\sum_i^{N_{\rm sub}}M_{t,i}\delta(\vec{x}-\vec{x}_i)\,.
\een
Neglecting sub-subhalos formally implies that $\vec{x}_i\neq \vec{x}_j\,\forall\,i\neq j$.
If we now take the continuous limit and make spherical symmetry explicit,
we obtain
\ben
\rho_{\rm tot}^2(R) &=& \rho_{\rm sm}^2(R)\nn\\
&& + \frac{N_{\rm sub}}{K_w}\,\frac{d{\cal P}_V(R)}{dV}\,
\int_{m_{\rm min}}^{m_{\rm max}}dm\,\frac{d{\cal P}_m(m)}{dm}\int_{c_{\rm min}(R)}^\infty dc\,
\frac{d{\cal P}_c(c,m)}{dc}\,\int_{0}^{r_t(R,c,m)} dr\,4\,\pi\,r^2\,
\left\{\rho^2(r) + 2\,\rho(r)\,\rho_{\rm sm}(R)\right\}\nn\\
&=& \rho_{\rm sm}^2(R)+
N_{\rm sub}\,\frac{d{\cal P}_V(R)}{dV}\,
\left\{ \rho^2_\odot \widetilde{\langle \xi_t \rangle}(R)
+ 2\,\rho_{\rm sm}(R)\,\widetilde{\langle m_t \rangle}(R) \right\}
\label{eq:rhotot2}
\een
where $m_t$ and $\xi_t$ are the subhalo tidal mass and annihilation volume, defined
in \citeeqss{eq:msub}{eq:xi} respectively. The symbol $\widetilde{\langle \rangle}$ denotes
the averaging over the concentration and mass parts of the subhalo phase space, made
explicit in \citeeq{eq:def_rhosub}, which is position dependent. Notice the crossing term
above induced by the interaction between subhalos and the host halo. Usually assumed to be
subleading and thereby neglected, it may actually dominate over the smooth contribution at large
Galactic radii, as will be shown below.
\end{widetext}
We now define the dimensionless WIMP {\em luminosity} ${\cal L}$, which measures the spatial
dependence of the annihilation rate, as
\ben
\label{eq:lum}
{\cal L}(R) \equiv \frac{\rho_{\rm tot}^2(R)}{\rho^2_\odot}\,,
\een
where the normalization $\rho_\odot$ is made at the solar position. We further introduce the
differential annihilation boost factor ${\cal B}(R)$, and the integrated annihilation boost factor
$B(R)$, as
\ben
\label{eq:boost}
{\cal B}(R) &\equiv& \frac{{\cal L}(R) }{{\cal L}_\text{no sub.}(R)}\,,\\
B(R) &=& \frac{\int_0^R dr\,r^2\,{\cal L}(r)}{\int_0^R dr\,r^2\,{\cal L}_\text{no sub.}(r)}\,.\nn
\een
Defined so, these boost factors are merely the multiplicative corrections to apply
to the differential or integrated annihilation rate as computed by neglecting the subhalo component.
Note that in principle, any WIMP signal prediction involving subhalos should be affected by a
statistical variance, reflecting the possible fluctuations of the number of contributing
objects \cite{Lavalle2007a,Lavalle2008c}. We keep this aspect for further dedicated studies.

We report our calculation results for the annihilation profiles in \citefig{fig:annprofiles}, where
we again adopt M11 as the reference Galactic mass model --- this is the translation of
\citefig{fig:mprofiles} in terms of annihilation profiles. Global tides are calculated from the
smooth Jacobi method [see \citeeq{eq:rt}], while disk-shocking effects are described from the
differential method [see \citeeq{eq:rt_diffds_it}]. The top (bottom) row's panels correspond to a
subhalo mass index of 1.9 (2). The left (right) column's panels correspond to a
minimal subhalo mass of $10^{-6}M_\odot$ ($10^{-10}M_\odot$). In each panel, the upper
part shows the different components of the annihilation profile, and the lower part shows the
differential and integrated boost factors as defined above. We display the impact of neglecting disk
shocking as dashed curves, which demonstrates the importance of this effect in the central parts
of the Galaxy.

A generic result is that the subhalo contribution dominates at large radii, typically from the
edge of the Galactic disk for a subhalo mass index of $\alpha_M=1.9$, and even from much more inner
regions in the case of $\alpha_M=2$. For an overall NFW profile, this leads to a characteristic
scaling of $1/r^2$ toward the Galactic center, where the smooth halo dominates, progressively
changing to $1/r^3$ outward, when the luminosity profile tracks the subhalo spatial
distribution. Interestingly, in the case of $\alpha_M=2$, where the global subhalo luminosity
is enhanced, a plateau arises in the overall luminosity profile at the transition
between smooth-halo domination and subhalo domination. This is actually an imprint of
disk-shocking effects, which delay the rising of the subhalo contribution. We will see later
that this plateau does not depend on the tidal disruption efficiency, and is also preserved
in the case of an overall Einasto profile. This striking feature might be used in gamma-ray
searches.

Finally, we comment on our results for the differential and integrated boost factors. The so-called
differential boost factor (reported as ``local'' in the plots) is mostly relevant to indirect dark
matter searches with antimatter cosmic
rays because of the limited horizon of the latter induced by propagation effects. It also
represents the correction to apply to the integrand of the line-of-sight integral used in
gamma-ray searches. On the other hand, the integrated boost is related to the absolute
Galactic luminosity, and thereby also to extragalactic gamma-ray searches --- then the Galaxy
appears as a template case characterizing other galaxies close in mass. We see that at the
solar position ($R/R_{200}\sim 0.03$-0.04), the boost is locally $<2$ for $\alpha_M=1.9$, while
it reaches $\sim 5$ for $\alpha_M=2$. Though moderate, these values may have some impact on
the existing limits on WIMPs as the precision in the cosmic-ray data has strongly increased
in recent years. We also remark that the differential boost increases up to $10^3$-$10^4$ toward
the edge of the Galaxy, which strongly affects, for instance, the diffuse gamma-ray signal at
high Galactic latitudes, as known from long ago
(see \eg\ Refs.~\cite{Bergstroem1999a,Pieri2011}). Regarding the integrated boost, the values
obtained at the edge of the dark halo can represent useful calibration values for calculations
of the dark matter contribution to the extragalactic diffuse gamma-ray flux. These go from
$\sim 3$ for $\alpha_M=1.9$, to $\sim 20$ for $\alpha_M=2$. This is fully consistent with the
recent study in Ref.~\cite{Moline2016} (see Fig. 6 in this article), which is based on fits of
cosmological simulations, and does not include baryonic effects -- while global tides are
merely the outcomes of the simulations themselves. That baryons play no role in the integrated
boost at the whole Galactic scale should not come as a surprise, as they only affect the
dynamics in the very central parts of the halo.

At this stage, we have illustrated our results assuming a tidal efficiency of $\varepsilon_t=1$
[see \citeeq{eq:disruption}]. It is important to check their stability against changes
in this parameter. In \citefig{fig:eps_impact}, we investigate the impact of $\varepsilon_t$
by computing the annihilation profiles for $\varepsilon_t=0.5$ (subhalos can be pruned down
to $r_s/2$ before getting disrupted), and for $\varepsilon_t=2$ (subhalos can be pruned only
down to $2\times r_s$ before getting disrupted). We adopt the configuration for which the
plateau discussed above is visible, namely $\alpha_M=2$. We see that the plateau is slightly
smeared when $\varepsilon_t=0.5$, as disk-shocking effects are then less disruptive and smear
the previously abrupt rising of the subhalo contribution. On the contrary, the plateau is
much more salient when $\varepsilon_t=2$, as expected. Values of 0.5 are rather small compared
to what is found in cosmological simulations (see \eg\ Ref.~\cite{Hayashi2003}), but they could
be relevant, for instance, to the very concentrated cores of ultracompact minihalos, close
to the minimal cutoff mass (see \eg\ Ref.~\cite{Berezinsky2014} for further discussion).
Nevertheless, we see that except close to the smooth-halo/subhalo luminosity domination
transition, changes in $\varepsilon_t$ have no significant impact on our predictions. In
particular, the plateau feature does not seem to be spoiled.
\begin{figure}[!t]
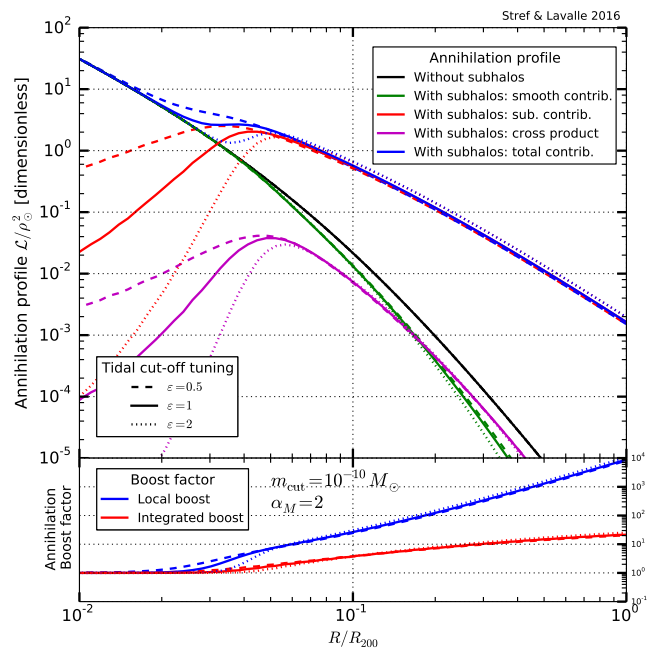

\centering
\includegraphics[width = 0.495\textwidth]{{{fig_annihilation_profile_smoothmod16_mmin1e-10_alpha2_comp_eps}}}
\caption{\small Impact of the tidal disruption efficiency on the annihilation profile
  [see \citeeq{eq:disruption}]. Decreasing
  values of $\varepsilon$ imply a less efficient disruption (tidal pruning allowed down to smaller
  radii).}
\label{fig:eps_impact}
\end{figure}
\subsection{Comparison between different Galactic mass models}
\label{ssec:comp_mass_models}
\begin{figure*}[!t]
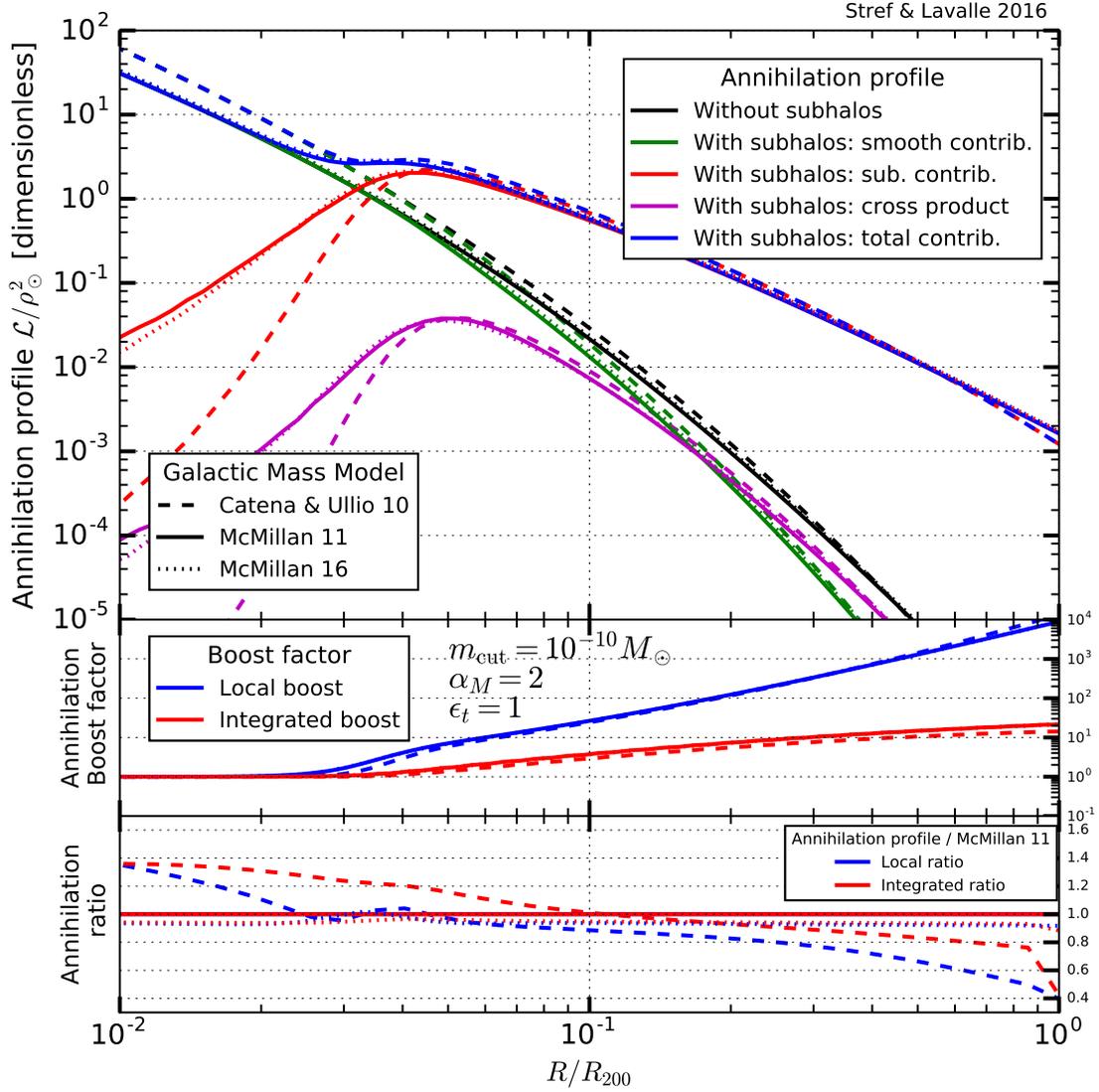

\centering
\includegraphics[width = 0.85\textwidth]{{{fig_comp_annihilation_profiles_mmin1e-10_alpha2_eps1}}}
\caption{\small Comparison of the annihilation profiles for different Galactic mass models
  (M11, CU10, and M16 -- see text for details). Annihilation profiles are normalized to
  the $\rho_\odot^2$ of each model. The lower panel shows the {\em real} relative difference of each
  configuration with the M11 setup.}
\label{fig:comp_mass_models}
\end{figure*}
Throughout this study, we have adopted the M11 Galactic mass model as a reference. It is
interesting to check how our predictions are affected by changes in the Galactic model
itself, while still trying to use dynamically constrained scenarios. To this aim, we
will use two other Galactic models: (i) that of Ref.~\cite{Catena2010a} (CU10 hereafter),
which is particularly interesting as it relies on an Einasto dark halo instead of an NFW one,
and (ii) the upgraded and updated version of M11, recently released in Ref.~\cite{McMillan2016}
by the same author (M16 henceforth), still based on an NFW profile for the dark halo,
but with additional constraints from novel kinematic data.
Besides the details of the dark matter halo profile and the data sets used as dynamical constraints,
changes with respect to M11 also come from the additional inclusion of a two-component gaseous
disk (HI and HII) in both models. Including gas actually has no significant impact on the dark
matter profile on large scales,
as this mostly tunes the distribution of the inner overall gravitational potential among all,
baryonic and dark, components. Rather, this mostly influences the understanding of very local
stellar dynamics. Note that since the gas components of CU10 are directly inferred from data points,
we will instead, for convenience, use those of M16 in both M16 and CU10, which may be described from
\citeeq{eq:rho_baryons}. (We further trade the original vertical sech$^2$ of M16 for an exponential
function, which has no impact on the final result). The sets of parameters of these models
are given in \citetab{tab:dark_halo} and \citetab{tab:baryons}.

Concrete comparisons are displayed in \citefig{fig:comp_mass_models} in terms of annihilation and
boost-factor profiles, where solid (dashed and dotted) curves correspond to M11 (CU10 and M16,
respectively) predictions. We adopt the ``best-case'' configuration, where $\alpha_M=2$ and
the cutoff subhalo mass is $10^{-10}M_\odot$. The color code is the same as in
\citefig{fig:annprofiles}. Luminosity profiles are measured in units of the squared dark matter
density $\rho_\odot^2$ at the solar position $R_\odot$, which (barely) change from one model to
another. The upper horizontal axes feature the Galactic radius in units of $R_{200}$, which also
varies between configurations -- see \citetab{tab:dark_halo}.
We add another lower panel that provides the real annihilation profile ratio with respect to M11,
where the luminosity is then evaluated at the corresponding M11 radius for each model --
consequently, the lower horizontal axis features $R/R_{200}$ as inferred from M11 only.

We first notice that the difference between M11 and M16 is hardly visible, and amounts to
$\lesssim 10\%$ over the full Galaxy, M16 being slightly less luminous in terms of dark matter
annihilation. This is not surprising as the only changes between M11 and M16 are the addition
of a gaseous disk and a new set of constraining data. On the other hand, differences are more
pronounced between CU10 and M11-M16: CU10 is brighter than M11-M16 in the central parts of the
Galaxy, typically for $R/R_{200}<0.1$. This is obviously a consequence of the different halo
shape, as Einasto profiles are known to be more luminous than NFW profiles within the scale
radius (except for the divergence of NFW profiles at the very centers -- see \eg\
Ref.~\cite{Pieri2011}), while having a faster luminosity decrease outward because of the
exponential cutoff in the halo shape. This amounts to an increase of $\sim 20\%$ in integrated
luminosity at the Solar position, and a decrease of the same order at the edge of the Galaxy.
In terms of boost factors, which measure the impact of subhalos relative to the
host halo and therefore can be directly compared between different mass models, we see that
the difference is very moderate for the differential boost, leading to an integrated difference $<2$
at the edge of the dark halo (CU10 leads to a slightly smaller integrated boost factor).

Finally, we remark that the plateau feature emphasized in \citesec{ssec:boosts} as a signature
of a sharp subhalo mass function also shows
up in CU10, despite the different overall halo shape. This prediction is therefore robust
against systematic uncertainties in the overall dark halo modeling, provided the smooth
halo density continues increasing inward, where the subhalo population has been fully depleted
--- this plateau could convert into a bump for a cored smooth halo profile.
\subsection{Impact of the minimal cutoff mass}
\label{ssec:mmin}
  As mentioned in \citesec{sssec:dPdm}, the minimal cutoff mass $m_{\rm min}$ of subhalos should be
  related to the very properties of the DM particle (production mechanism, interaction properties).
  In principle, one can calculate this cutoff mass for each candidate \cite{Chen2001,Hofmann2001,Berezinsky2003,Green2004a,Loeb2005,Boehm2005a,Profumo2006,Bertschinger2006a,Bringmann2007,Bringmann2009,Berezinsky2014,Visinelli2015}, so it is not, strictly speaking, a free parameter once the
  underlying theory is fixed, even though equal-mass particles would come with minimal mass scale
  varying by orders of magnitude \cite{Bringmann2009}. Nevertheless, it is still interesting to
  summarize the impact of $m_{\rm min}$ on our results. The following discussion assumes that
  the initial spectral mass index is fixed to a single value $\alpha_M>1$, so that the structure
  mass
  spectrum is regular -- note that some inflation or phase transition scenarios may predict more
  complex patterns for the primordial perturbation spectrum on small scales which should imprint
  the mass spectrum, strongly affecting the way we described the initial mass function
  (see \eg~\cite{Ricotti2009} and further exploration of this framework in
  \cite{AslanyanEtAl2016}); we will not discuss these cases here.

  Since we have calibrated the subhalo population by fixing a mass fraction in a reference
  mass range (see \citesec{sssec:norm_sub}), and since the maximal subhalo mass is
  $m_{\rm max}\gg m_{\rm min}$ it is easy to show that the total number of subhalos scales like
  $N_{\rm sub}\overset{\sim}{\propto} m_{\rm min}^{1-\alpha_M}$, and hence the number density profiles
  $n_{\rm sub}(r)$ shown in \citefig{fig:nprofile}.
  Tidal effects will simply affect the relative
  fraction of the less massive objects with respect to the more massive ones toward the central
  regions of the Galaxy, as shown in \citefig{fig:dndm}. This is reflected by the local modification
  of the initial mass index, as discussed below \citeeq{eq:dpeffdm}.

  For the mass profiles
  shown in \citefig{fig:mprofiles}, the impact of $m_{\rm min}$ can be inferred from the relation
  $\rho_{\rm sub}\overset{\sim}{\propto} N_{\rm sub}\,\langle m \rangle \approx {\rm cst}$,
  with $\langle m \rangle \overset{\sim}{\propto} m_{\rm min}^{\alpha_M-1}\,\ln(m_{\rm max}/m_{\rm min})$
  (we have assumed $\alpha_M\sim 2$ to get the log). Given the scaling of $N_{\rm sub}$ discussed
  just above, this helps understand that despite the striking difference in terms on number
  density, the mass profile is instead rather similar for $m_{\rm min}=10^{-6}\,M_\odot$ and
  $m_{\rm min}=10^{-10}\,M_\odot$. Indeed, the small difference induced by $m_{\rm min}$ on
  $\rho_{\rm sub}$ comes from the logarithmic correction (a factor of 4/5) and also from tidal
  stripping which is more efficient for more massive subhalos,
  of which the relative population is further reduced in the latter case, hence a slightly larger
  amplitude for the corresponding mass profiles. This difference is a bit more pronounced for
  $\alpha_M=2$ because of the logarithmic term discussed above, which is replaced by
  $\sim m_{\rm max}^{2-\alpha_M}$ for $\alpha_M=1.9$, further reducing the impact of $m_{\rm min}$.

  Finally, the dependence of the annihilation profiles depicted in \citefig{fig:annprofiles}
  in $m_{\rm min}$ can be understood from the subhalo mass dependence of its annihilation
  luminosity, which for an NFW profile is ${\cal L}_i \overset{\sim}{\propto} m_i^{0.9}$
  \cite{Lavalle2008c}. Since the total luminosity ${\cal L} \overset{\sim}{\propto} N_{\rm sub}
  \, \langle {\cal L}_i \rangle $, then we get
  ${\cal L} \overset{\sim}{\propto} m_{\rm min}^{1.9-\alpha_M}$, which explains why the annihilation
  profiles do not exhibit significant dependence on $m_{\rm min}$ when $\alpha_M=1.9$, while it is
  more noticeable for $\alpha_M=2$.
\section{Conclusions}
\label{sec:concl}
We have proposed a method to model a galactic dark matter subhalo population consistently with
dynamical constraints, focusing on the Milky Way -- subhalos are unavoidable galactic components
if dark matter is made of WIMPs or any other dark matter candidates with suppressed
self-interactions and devoid of additional pressure. Dynamical consistency is important to make
sense of constraints or discovery potentials of both direct and indirect dark matter searches
(see \eg\ Ref.~\cite{Lavalle2015} for an illustration in direct searches). We have assumed that
subhalos initially track
the host halo profile when the galaxy forms and then we have explicitly calculated the effects
of tidal stripping and subsequent potential disruption induced not only by the overall
gravitational potential but also by baryonic disk crossing. We have developed and compared
different theoretical approaches to deal with the latter, and retained the so-called differential
disk-shocking method as our reference case, since it is built upon more accurate physical grounds.
This method was inspired by previous works dedicated to the understanding of stellar
clusters, in particular by
Refs.~\cite{Ostriker1972,Weinberg1994,Gnedin1999,Gnedin1999a,Gnedin1999b}. These references
were already used in other analytic studies (see \eg\ Ref.~\cite{Berezinsky2008}), but were
dealt with in a significantly different way, leading to a different formulation of tidal mass
losses without explicit links to the definition of a tidal radius. Nevertheless, even if it is
difficult to make quantitative comparisons between the mentioned study and ours, it seems that
both approaches are in agreement at least at the qualitative level. Our study was more aimed at
quantifying the impact of a subhalo population on the dark matter annihilation rate and sizing the
related theoretical uncertainties in a realistic Galactic mass configuration, while
Ref.~\cite{Berezinsky2008} was more concerned with the survival probability of subhalos against
different types of tidal effects. On the whole, our method to include disk-shocking effects is
likely simpler to implement in numerical calculations. A summary of the method
is given in \citesec{sec:sketch}.

The main inputs of our model are (i) the Galactic mass model, (ii) the subhalo mass function,
and (iii) the subhalo concentration function, for which we adopted consensual prescriptions.
Further assumptions regard the choice of the
inner subhalo profile. We considered a spherically symmetric host halo, hence a spherically
symmetric subhalo distribution. Our model can in principle easily be extended to axisymmetric
host halos, while its numerical implementation will then likely become much trickier. 
We stress that we calibrated the subhalo mass fraction using constraints from cosmological
simulations {\em without} baryons. It is important to use dark-matter-only simulations
because baryonic components in hydrodynamic simulation are likely to differ significantly
from those of the real Milky Way, while strongly affecting the dynamics of subhalos: it would then
become impossible to disentangle the global from other tidal effects, and it would make it spurious
to calibrate a subhalo population model {\em a posteriori}. This underlines the need to
continue running dark-matter-only simulations with increased resolution and up-to-date
cosmological parameters, even in a context where issues related to the impact of baryons
on cold dark matter halos are certainly the most pressing ones.

Using the recent and constrained Galactic mass models from
Refs.~\cite{McMillan2011,Catena2010a,McMillan2016} (dubbed M11, CU10, and M16 -- M11 being used as
the template case), characterized by different assumptions on the dark halo profile while
providing results consistent with each other, we computed the overall subhalo mass profiles and
further made predictions for the induced annihilation profiles. We stress that these results
incorporate a self-consistent calculation of the tidal radius for each subhalo, depending on its
mass, concentration, and position in the Galaxy; individual mass and luminosity are calculated
up to the tidal radius for each subhalo. We used different assumptions for
the mass index $\alpha_M$ and the cutoff subhalo mass. Since the latter could in principle be
determined from WIMP interaction properties in specific scenarios, the main theoretical
uncertainty is actually carried by $\alpha_M$. Based on our reference model summarized
in \citesec{ssec:ref_mod}, we showed that the global or integrated boost
factor could vary between $\sim 2$ (for $\alpha_M=1.9$) and $\sim 20$ (for $\alpha_M=2$,
respectively) for all choices of Galactic mass models. This may provide interesting and
complementary calibration points for estimates of the dark matter contribution to the
extragalactic diffuse gamma-ray background (see \eg\ Ref.~\cite{Ajello2015}). We also derived
differential boost factors ({\em i.e.} the boost factor profile)
that could be used to revisit estimates of the dark matter contribution to the Galactic diffuse
gamma-ray emission (\eg\ Refs.~\cite{Charbonnier2012,Ackermann2012h,Ackermann2015,FornasaEtAl2016,Cirelli2015}),
or to the local antimatter cosmic-ray flux
(\eg\ Refs.~\cite{Boudaud2015a,Giesen2015,Kappl2015a,Evoli2015}). Interestingly, our model predicts
a plateau in the overall annihilation rate in the case of a sharp mass function ($\alpha_M=2$)
that could lead to specific observable effects. This feature seems to persist within the
considered theoretical and systematic uncertainties of the model.

The local subhalo population and induced boost factor, relevant to direct searches and antimatter
searches, respectively, are very sensitive to $\alpha_M$. For antimatter searches, though,
the precision achieved in the most recent measurements is such that even a moderate effect could
have a significant impact on the existing limits or discovery prospects. In the most optimistic
case, when $\alpha_M=2$, the enhancement can reach a factor of 10 (the impact of the
cutoff subhalo mass $m_{\rm min}$ is discussed in \citesec{ssec:mmin}).

It would be interesting to test this model against cosmological simulations with baryons in the
relevant subhalo mass range in the near future, but this is clearly beyond the scope of the
present study. In any case, the model is easily tuneable in terms of initial distribution functions,
provided internal consistency with the dynamical constraints, which was the main purpose of this
work. Finally, self-made predictions for direct and indirect dark matter searches are left to
further studies.

\acknowledgments{
  We would like to thank the anonymous referees for their constructive comments, which
  helped improve the presentation of our results.
  The PhD grant of MS is funded by the OCEVU Labex (ANR-11-LABX-0060), which also
  provided financial support to this project. We also acknowledge support from the CNRS program
  {\em D\'efi InPhyNiTi}, and from the European Union's Horizon 2020 research and innovation
  program under the Marie Sk\l{}odowska-Curie grant agreements No 690575 and  No 674896; beside
  recurrent institutional funding by CNRS and the University of Montpellier.
}

\bibliography{biblio_jabref}

\end{document}